\documentclass[aps, prd, 12pt, eqsecnum, tightenlines, notitlepage, superscriptaddress, nofootinbib, floatfix, preprintnumbers]{revtex4-1}
\pdfoutput=1
\PassOptionsToPackage{unicode}{hyperref}

\usepackage{hyperref, graphicx, multirow, slashed, xcolor, amsmath, breakurl}

\allowdisplaybreaks

\definecolor{red}{rgb}{1,0,0}

\newcommand{\TeV}{\text{TeV}}
\newcommand{\GeV}{\text{GeV}}
\newcommand{\MeV}{\text{MeV}}

\begin{document}

\preprint{CALT-TH 2015-029}
\preprint{KANAZAWA-15-07}

\title{New Vector-Like Fermions and Flavor Physics}

\author{Koji Ishiwata}
\affiliation{Institute for Theoretical Physics, Kanazawa University, Kanazawa
920-1192, Japan}

\author{Zoltan Ligeti}
\affiliation{Ernest Orlando Lawrence Berkeley National Laboratory,
University of California, Berkeley, CA 94720}

\author{Mark B.\ Wise}
\affiliation{Walter Burke Institute for Theoretical Physics,
California Institute of Technology, Pasadena, CA 91125}

\begin{abstract}

We study renormalizable extensions of the standard model that contain
vector-like fermions in a (single) complex representation of the
standard model gauge group.  There are 11 models where the vector-like
fermions Yukawa couple to the standard model fermions via the Higgs
field.  These models do not introduce additional fine-tunings.  They
can lead to, and are constrained by, a number of different
flavor-changing processes involving leptons and quarks, as well as
direct searches.  An interesting feature of the models with strongly
interacting vector-like fermions is that constraints from neutral
meson mixings (apart from $CP$ violation in $K^0- {\bar K}^0$ mixing)
are not sensitive to higher scales than other flavor-changing
neutral-current processes. We identify order $1/(4 \pi M)^2$ (where
$M$ is the vector-like fermion mass) one-loop contributions to the
coefficients of the four-quark operators for meson mixing, that are
not suppressed by standard model quark masses and/or mixing angles.

\end{abstract}

\maketitle

\section{Introduction}

The driving force behind many of the extensions of the standard model
(SM) has been the hierarchy puzzle. However, there may be reasons that
the fine tuning of quadratically large contributions to the Higgs mass
from very high momentum scales is acceptable (e.g., environmental
reasons).  This motivates the consideration of simple extensions of
the standard model that do not address the hierarchy puzzle and do not
introduce any additional fine tunings of parameters.

The SM provides no understanding why there are three generations of
chiral fermions with masses that require weak symmetry breaking or why
they only come in such simple representations of the gauge
group. Vector-like fermions can be much heavier than the SM fermions,
since their masses do not require weak symmetry breaking. For some
choices of quantum numbers, such vector-like fermions can Yukawa
couple to SM fermions.  Given our lack of understanding of the number
of generations and their quantum numbers, it seems worthwhile
exploring the possibility that vector-like fermions exist. Of course,
their masses may all be much larger than the weak scale, but it is not
unreasonable that one such vector-like representation has a mass light
enough that it can influence the next generation of flavor physics
experiments. This motivates a study of extensions of the standard
model with additional fermions that reside in a single vector-like
representation of the SM gauge group with a mass $M$.  (For some
earlier studies of such models, see, e.g., Refs.~\cite{Branco:1986my,
  Nir:1990yq, delAguila:1989rq, Barenboim:2001fd, Cacciapaglia:2010vn,
  Altmannshofer:2013zba, Ma:2013tda, Ellis:2014dza, delAguila:2000rc,
  Alok:2015iha}.)  We focus on vector-like fermions that can have
Yukawa couplings involving the Higgs field and the SM fermions.  This
allows them to influence flavor physics. There are several such
models. These models have been studied in the literature previously,
but here we consider all such models and compare the patterns of
deviations from the SM they would give rise to. These models predict a
very specific pattern for the contributions of beyond the standard
model (BSM) physics to $\Delta F=1$ flavor-changing neutral-current
(FCNC) processes compared to $\Delta F=2$ neutral meson mixings.  For
large masses $M$ the dominant order $1/(4 \pi M)^2$ contribution to
the coefficients of the four-quark operators responsible for neutral
meson mixing arise at one loop, and are not suppressed by SM quark
masses and/or weak mixing angles. We compute this contribution to
meson mixing in the 7 models that involve strongly interacting
vector-like fermions. In addition to considering $\Delta F=1$ and
$\Delta F=2$ flavor-changing neutral-current processes, we also
briefly discuss violations of lepton universality and unitarity of the
CKM matrix that arises from the corrections to the $W$ boson couplings
in these models.

The light neutrino masses play no role in
our analysis so we take the SM to contain massless left-handed
neutrinos. We are most interested in experiments that have reach in
vector-like fermion mass of more than $\sim$\,10\,TeV when the Yukawa
coupling constants of the vector-like fermions to the standard model
fermions are around unity.  For such masses, effects originating from
the non-unitarity of the $3 \times 3$ CKM matrix can be neglected.

If the new fermion is in a real representation it can have a Majorana
mass term which does not require a helicity partner in the
spectrum. In that case, unless the Majorana mass is very large or the
Yukawa coupling of the new neutral fermion to the SM neutrinos is very
small, a SM neutrino gets an unacceptably large mass. Hence we do not
consider models where the additional fermions come in a real
representation of the gauge group.\footnote{There are two cases of
this type. New fermions with SM quantum numbers $(1,1,0)$ and
$(1,3,0)$.}

ATLAS and CMS have searched for strongly interacting vector-like
fermions contained in several of the models discussed
below~\cite{Aad:2015mba, Succurro:2013doa, Khachatryan:2015axa,
  Chatrchyan:2013wfa, Chatrchyan:2013uxa}.  The bounds from the Run~1
data constrain the masses of vector-like quarks typically at about the
$M > 800\,\GeV$ level.  These constraints come from the pair
production of the vector-like fermions via their gauge couplings, and
are (essentially) independent of the Yukawa couplings, which are the
focus of this paper.

In Section~II we describe the models, Section~III explores the
constraints from measurements both in the lepton and quark sectors,
and Section~IV contains our conclusions.

\section{The models}

There are 11 renormalizable models with vector-like fermions in
complex representations of the standard model gauge group, where the
vector-like fermions have renormalizable Yukawa couplings to the SM
fermions through the Higgs doublet.

Models~I and II have $SU(2)_L$ singlet, $E$, and $SU(2)_L$ triplet,
$T_e$, vector-like fermions. Their $SU(3)_C \times SU(2)_L\times
U(1)_Y$ quantum numbers are $(1,1,-1)$ and $(1,3,-1)$. These
vector-like fermions Yukawa couple through the Higgs field to the SM
left-handed lepton doublets $L^i$.

Model~III has vector-like fermions, $\ell$, with same gauge quantum
numbers as the SM left-handed doublets, $(1,2,-1/2)$. They Yukawa
couple to the SM right-handed charged lepton fields $e_R^i$.
Model~IV contains vector-like fermions, $l$, with $(1,2,-3/2)$
gauge quantum numbers, and they also Yukawa couple to $e_R^i$.

Models V and VI contain vector-like fermions $D$ and $U$ with the same
quantum numbers as the SM right-handed down and up-type quarks,
$(3,1,-1/3)$ and $(3,1,2/3)$. They Yukawa couple to the SM left-handed
quark doublets $Q_L^i$. Models~VII and VIII are similar with
vector-like $SU(2)_L$ triplet fermions $T_d$ and $T_u$ with gauge
quantum numbers $(3,3,-1/3)$ and $(3,1,2/3)$ respectively.

There are three more models where the vector-like fermions Yukawa
couple to the SM right-handed up-type, $u_R^i$ and down-type $d_R^i$
quarks. In Model~IX the vector-like fermions $q_{ud}$ have the same
quantum numbers as the SM left-handed quark doublets, $(3,2,1/6)$. In
this case the new vector-like quark doublet couples to both $u_R^i$
and $d_R^i$. Model~X has a vector-like quark doublet $q_u$ with
quantum numbers $(3,2,7/6)$ that Yukawa couple through the Higgs
doublet to the right-handed up-type quarks.  Model~XI has a
vector-like quark doublet $q_d$ with quantum numbers $(3,2,-5/6)$ that
Yukawa couples to the right-handed down-type quarks.

\subsection{Lagrangians}

The new terms in  the Lagrange density which are added to the SM Lagrange
density are:
\begin{align}
{\rm Model~I}\ (1,1,-1):~
  &{\cal L }_{\rm BSM} = {\bar E} \left(
  i\slashed{D} - M \right)E - \left( \lambda_i{ \bar E_R} H^{\dagger} L^i_L 
  + h.c. \right) , \\
{\rm Model~II}\ (1,3,-1):~
  &{\cal L}_{\rm BSM} = 
  {\rm Tr}\left[ {\bar T_e} \left( i\slashed{D} - M \right)T_e \right] -
  \left( \lambda_i  H^{\dagger} {\bar T_{eR}} L_L^i+ h.c.\right) , \\  
{\rm Model~III}\ (1,2,-1/2):~
  &{\cal L}_{\rm BSM}={\bar \ell} \left( i\slashed{D} - M \right) \ell
  - \left( \lambda_i \bar{e}_R^i H^{\dagger} \ell_L + h.c.\right) , \\
{\rm Model~IV}\ (1,2,-3/2):~
  &{\cal L}_{\rm BSM}={\bar l} \left( i\slashed{D} - M \right) l
  - \left( \lambda_i \bar{e}_R^i H^T \epsilon\, l_L + h.c.\right) , \\
{\rm Model~V}\ (3,1,-1/3):~
  &{\cal  L }_{\rm BSM} = {\bar D} \left( i\slashed{D} - M \right)D -
  \left(\lambda_i {\bar D_R} H^{\dagger} Q_L^i + h.c.\right) , \\
{\rm Model~VI}\ (3,1,2/3):~
  &{\cal L}_{\rm BSM} = {\bar U} \left( i\slashed{D} - M \right)U -
  \left(\lambda_i {\bar U_R} H^T \epsilon\, Q_L^i + h.c.\right) , \\
{\rm Model~VII}\ (3,3,-1/3):~
  &{\cal L}_{\rm BSM} = {\rm Tr}\left[ {\bar T_d}
  \left( i\slashed{D} - M \right)T_d \right] -
  \left(\lambda_i H^{\dagger} {\bar T_{dR}}Q_L^i + h.c.\right) , \\
{\rm Model~VIII}\ (3,3,2/3):~
  &{\cal L}_{\rm BSM} = {\rm Tr}\left[ {\bar T_u}
  \left( i\slashed{D} - M \right)T_u \right]
  - \left(\lambda_i H^{T}\epsilon\, \bar T_{uR}\, Q_L^i + h.c.\right) , \\
{\rm Model~IX}\ (3,2,1/6):~
  &{\cal L}_{\rm BSM} = {\bar q}_{ud} \left( i\slashed{D} - M \right)q_{ud}
  - \big(\lambda^{(u)}_i {\bar u_R}^i H^T \epsilon\, q_{udL}
  +\lambda^{(d)}_i {\bar d_R}^i H^{\dagger}q_{udL} + h.c.\big) , \\
{\rm Model~X}\ (3,2,7/6):~
  &{\cal  L }_{\rm BSM}={\bar q_u} \left( i\slashed{D}
  - M \right)q_u - \left(\lambda_i {\bar u_R}^i H^{\dagger}
  q_{uL} + h.c.\right) , \\
{\rm Model~XI}\ (3,2,-5/6):~
  &{\cal L}_{\rm BSM}={\bar q_d} \left( i\slashed{D}
  - M \right)q_d - \left(\lambda_i {\bar d_R}^iH^{T}\epsilon\, q_{dL}
  + h.c.\right) .
\end{align}
Here the $SU(2)$ triplets $T_e$, $T_u$ and $T_d$ are represented by
the two-by-two matrices
\begin{align}
T_e &= \left( \begin{array}{cc}
  T_{-1}/\sqrt{2} & T_0  \\
  T_{-2} &-T_{-1}/\sqrt{2} \end{array} \right) , \qquad\quad\ \,
  {\bar T}_e=\left( \begin{array}{cc}
  {\bar T}_{-1}/\sqrt{2} & {\bar T}_{-2}  \\
  {\bar T}_{0} &-{\bar T}_{-1}/\sqrt{2} \end{array} \right) , \\
T_d &= \left( \begin{array}{cc}
  T_{-1/3}/\sqrt{2} & T_{2/3}  \\
  T_{-4/3} &-T_{-1/3}/\sqrt{2} \end{array} \right) , \qquad
  {\bar T}_d = \left( \begin{array}{cc}
  {\bar T}_{-1/3}/\sqrt{2} & {\bar T}_{-4/3} \\
  {\bar T}_{2/3} &-{\bar T}_{-1/3}/\sqrt{2} \end{array} \right) , \\
T_u &= \left( \begin{array}{cc}
T_{2/3}/\sqrt{2} & T_{5/3}  \\
T_{-1/3} &-T_{2/3}/\sqrt{2} \end{array} \right) , \qquad\quad
{\bar T}_u=\left( \begin{array}{cc}
{\bar T}_{2/3}/\sqrt{2} & {\bar T}_{-1/3}  \\
{\bar T}_{5/3} &-{\bar T}_{2/3}/\sqrt{2} \end{array} \right) ,
\end{align} 
where the subscripts denote the charges of the fermions.

It is important to note that there are no new one-loop contributions
to the ordinary lepton and quark mass matrices of the form
\begin{equation}
m_{f_{ij}} \sim {\lambda_{i}^*\lambda_j\, v \over 16 \pi^2} \, ,
\end{equation}
where $v\simeq 174~{\rm GeV}$ is the Higgs vacuum expectation
value. This is due to the approximate chiral symmetry, e.g., $e_R^i
\rightarrow e^{i \alpha}\, e_R^i$ in Model~I, which is only broken by
the SM Yukawa couplings.  This prevents such contributions to the
light fermion mass matrix. So the $\lambda^i$ couplings can be order
unity without the need for a cancellation in the SM fermion mass
matrix.

The BSM effects caused by the vector-like fermions vanish as their
mass $M \rightarrow \infty$. We are interested in very large $M$,
greater than 10~TeV, and it is flavor-changing neutral-current
processes that are sensitive to such BSM physics. In the SM,
flavor-changing neutral currents are suppressed by small weak mixing
angles and small SM quark or lepton masses. Furthermore, they do not
occur at tree level.  Since the neutrino masses are very small,
charged lepton flavor violation in the SM is negligibly small.  In the
quark sector, despite their suppression, flavor-changing neutral-current 
processes that change flavor by one ($\Delta F=1$, e.g., $K^+
\rightarrow \pi^+ \nu {\bar \nu} $) and by two ($\Delta F=2$, e.g.,
$K^0-{\bar K}^0$ mixing) have been observed.

In the limit where the SM Yukawa couplings vanish, the SM has a
$U(3)_Q\times U(3)_{u_R} \times U(3)_{d_R} \times U(3)_L \times
U(3)_{e_R}$ flavor symmetry. This symmetry forbids flavor-changing
neutral currents. However in Models~I--XI, even when the SM quark and
lepton Yukawas are zero, the terms in the Lagrange density
proportional to the Yukawa couplings $\lambda_i$ break that
symmetry. That pattern of flavor symmetry breaking can be
characterized using the spurion method. For example in Model~V the BSM
terms in the Lagrange density involving $\lambda_i$ are not invariant
under the $U(3)_Q$ transformations $Q_L^i \rightarrow V(Q_L)_{ij}\,
Q_L^j$ (repeated flavor indices are summed).  However, if the
$\lambda_i$ also transform as $\lambda_i \rightarrow V(Q_L)^{*}_{ij}
\lambda_j$, then the $U(3)_{Q}$ symmetry is restored. This means that
in Model~V one-loop BSM physics associated with the high mass scale
$M$ generates meson mixing through a term in the effective Lagrangian
of the form,
\begin{equation}
{\cal L}_{\rm meson}^{\rm (V)} \sim 
  {\lambda_i^* \lambda_j \lambda_k^* \lambda_l \over (4 \pi M)^2}\,
  ({\bar  Q^i}_L\gamma_{\mu} Q^j_L)\, ({\bar  Q^k}_L\gamma^{\mu} Q^l_L)\,,
\end{equation}
and similarly for the other models. These corrections are not
suppressed by small quark masses and/or mixing angles, and we have not
found expressions for them in the literature. (We take the BSM Yukawa
couplings $\lambda_i$ to be of order unity.)  They are computed for
Models~V--XI in Sec.~\ref{sec:mix}.  Corrections to meson mixing of
order $\lambda_i \lambda_j^*/(4\pi M)^2$ that are also suppressed by
weak mixing angles and/or quark masses were considered for Model~V in
Ref.~\cite{Barenboim:1997pf}.

\subsection{New interactions with gauge bosons}

The new BSM contribution to the $Z$ coupling and to the $W$ arise at
tree level, after integrating out the heavy fermion(s). They can be
obtained ether by calculating Feynman diagrams or by diagonalizing the
$4\times 4$ fermion mass matrices as was discussed in
Ref.~\cite{Ishiwata:2013gma}. Here we explicitly show how to obtain
the BSM contributions to the $Z$ couplings for Model~I. (The same
method can be applied to the other models.)

After electroweak symmetry breaking, the mass matrix in charged lepton
sector becomes,
\begin{equation}
  {\cal L}_{m^{\hat{e}}}=
  -\bar{{\hat e}}_L^{A} {\cal M }_{AB}^{\hat{e}} {\hat e}_R^B +h.c.\, ,
\end{equation}
where roman capital indices $A$ and $B$ go over $\{0,1,2,3\}$,
${\hat e}$ is defined by ${\hat e}\equiv (E, e^1, e^2, e^3)^T$, and
\begin{equation}
{\cal M}^{\hat{e}} = \left( \begin{array}{cccc}
  M & 0&0&0  \\
  \lambda_1^* v &m_{e1} & 0&0  \\
 \lambda_2^* v &0&m_{e2} & 0  \\
 \lambda_3^* v &0 & 0 & m_{e3} 
\end{array} \right).
\end{equation}
Here we assumed, without loss of generality, that the charged lepton
fields $e_L^i$ (i.e., the lower components of the doublets $L_L^i$)
are eigenstates of the charged lepton mass matrix in the SM (i.e., in
the $\lambda_i\to 0$ limit).  This matrix is diagonalized by the
$4\times 4$ unitary transformations $V_{L,R}^{\hat{e}}$,
\begin{equation}
{\hat e}'_{L,R}=V_{L,R}^{\hat{e}}\, {\hat e}_{L,R}\, ,
\end{equation} 
where the prime denotes a mass eigenstate field. So
\begin{equation}
V_L^{\hat{e}} {\cal M}^{\hat{e}}V_R^{\hat{e}\dagger}={\cal M}_{\rm diag}^{\hat{e}}\,.
\end{equation}
Up to corrections suppressed by $(v/M)^2$
\begin{equation}
V_L^{\hat{e}} = \left( \begin{array}{cccc}
1 & \lambda_1 v/M& \lambda_2 v/M & \lambda_3 v/M  \\
-\lambda_1^* v/M &1 & 0&0  \\
-\lambda_2^* v/M &0&1 & 0  \\
-\lambda_3^* v/M &0 & 0 & 1 
\end{array} \right), \qquad 
  V_R^{\hat{e}} = 1_{4\times 4} \,.
\end{equation}
Consequently the masses of the charged leptons in the SM part of the
Lagrangian are approximately equal to the charged lepton masses, and
the vector mass parameter $M$ is approximately the heavy vector-like
lepton mass.

The $Z$ boson coupling is,
\begin{equation}
{\cal L}^{(Z)} = -g_Z Z_{\mu} \left[ \frac12\, {\bar{\hat e}}{}_L^{\prime A}\,
  (V_L^{\hat{e}})_{A0}\, \gamma^\mu (V_L^{\hat{e}\dagger})_{0B}\,
  {\hat e}_L^{\prime B} + \ldots \right] ,
\end{equation}
where the ellipses denote terms not containing the matrix $V_L^{\hat
  e}$.  Here $g_Z=\sqrt{g_1^2+g_2^2}$ and $g_{1,2}$ are the gauge
couplings of $U(1)_Y$ and $SU(2)_L$, respectively.  There is some
ambiguity in how the terms are organized since $V_L^{\hat e}$ is
unitary. We have written the $Z$ couplings involving $V_L^{\hat e}$ so
that the order $(v/M)^2$ corrections to the SM $Z$ couplings to the
charged leptons can be read off using the order $(v/M)$ terms in
$V_L^{\hat{e}}$ that we have explicitly calculated. From this it
follows (removing the hats and primes) that the BSM couplings of the
$Z$ boson to the light mass eigenstate charged leptons are,
\begin{eqnarray}
\label{newzcoup1}
{\rm Model~I}:~{\cal L}_{\rm BSM}^{(Z)} =
-\sum_{i,j}
\left( {\lambda_i^* \lambda_j m_Z^2 \over g_Z M^2}\right) {\bar e}^{i}_L
\gamma^{\mu} e^{j}_L Z_{\mu} \, ,
\end{eqnarray}
with $m_Z$ being the $Z$ boson mass. For simplicity, hereafter we
remove the primes used to specify the mass eigenstate fields.

In the same way, the new $Z$ couplings to the quarks and charged
leptons are obtained for the other models. (The $4\times 4$ unitary
matrices are given in Appendix~\ref{app:DiagMatrix}.)  The results are
\begin{align}
{\rm Model~II}:~&{\cal L}_{\rm BSM}^{(Z)}= -\sum_{i,j}
  \left( {\lambda_i ^*\lambda_j m_Z^2 \over 2g_ZM^2}\right) {\bar e}^{i}_L 
  \gamma^{\mu} e^{j}_L\, Z_{\mu} \,, \label{newzcoup2} \\
{\rm Model~III}:~&{\cal L}_{\rm BSM}^{(Z)} = \sum_{i,j}  
  \left( {\lambda_i  \lambda_j^* m_Z^2 \over g_Z M^2}\right)
     {\bar e}^{i}_R \gamma^{\mu} e^{j}_R\, Z_{\mu} \,, \label{newzcoup3} \\   
{\rm Model~IV}:~&{\cal L}_{\rm BSM}^{(Z)} = - \sum_{i,j}  
  \left( {\lambda_i  \lambda_j^* m_Z^2 \over g_Z M^2}\right)
     {\bar e}^{i}_R \gamma^{\mu} e^{j}_R\, Z_{\mu} \, .
     \label{newzcoup4}
\end{align}
Note that in Models~III and IV the BSM tree-level $Z$ couplings differ only by
an overall sign.

Similarly for hadronic models,
\begin{align}
{\rm Model~V}:~
&{\cal L}_{\rm BSM}^{(Z)}=
-\sum_{i, j} \left( {\lambda_i^* \lambda_j m_Z^2 \over g_Z M^2}\right) 
  {\bar d}^{i}_L \gamma^{\mu} d^{j}_L\, Z_{\mu} \,, \label{newzcoup5} \\
{\rm Model~VI}:~
&{\cal L}_{\rm BSM}^{(Z)}= \sum_{i, j}
\left( {\lambda_i^* \lambda_j m_Z^2 \over g_Z M^2}\right) 
  {\bar u}^{i}_L \gamma^{\mu} u^{j}_L\, Z_{\mu} \,, \label{newzcoup6} \\
{\rm Model~IX}:~
&{\cal L}_{\rm BSM}^{(Z)}= -\sum_{i,j} \Bigg( 
{\lambda_i^{(u)} \lambda_j^{(u)*} m_Z^2 \over g_ZM^2}\, 
  {\bar u}^{i}_R \gamma^{\mu} u^{j}_R 
- {\lambda_i^{(d)} \lambda_j^{(d)*} m_Z^2 \over g_Z M^2}\, {\bar d}^{i}_R
\gamma^{\alpha} d^{j}_R \Bigg) Z_{\mu} \,, \label{newzcoup9}  \\
{\rm Model~X}:~
&{\cal L}_{\rm BSM}^{(Z)}=
\sum_{i,j} \left( {\lambda_i \lambda_j^* m_Z^2 \over g_Z M^2}\right)
  {\bar u}^{i}_R \gamma^{\mu} u^{j}_R\, Z_{\mu} \,, \label{newzcoup10} \\
{\rm Model~XI}:~
&{\cal L}_{\rm BSM}^{(Z)}=
  -\sum_{i,j} \left( {\lambda_i \lambda_j^* m_Z^2 \over g_Z M^2}\right)
  {\bar d}^{i}_R \gamma^{\mu} d^{j}_R\, Z_{\mu} \,. \label{newzcoup11}
\end{align}

So far, we have worked in the mass eigenstate basis most convenient
for expressing the $Z$ couplings. For example, in Model~V we have
worked in a basis where in the $\lambda_i\to 0$ limit the down-type
quarks in the Lagrangian are mass eigenstates, while in Model~VI we
have worked in the basis where the down-type quarks in the Lagrangian
are mass eigenstate. In Model~IX the couplings $\lambda^{(d)}$
correspond to the down quark mass eigenstate basis and the couplings
$\lambda^{(u)}$ are in the up-quark mass eigenstate basis.

In case of vector-like triplets, i.e., Models~VII and VIII, it is not
possible to choose a basis where the CKM matrix, $V$, is absent from
the $Z$ couplings.\footnote{More precisely, the unitary matrix $V$ is
  the CKM matrix in the $\lambda_i\to 0$ limit.}  Choosing the up-type
quarks in the Lagrange density to be mass eigenstates (in the
$\lambda_i \to 0$ limit) implies for Model~VII,
\begin{equation}
\label{newzcoup7}
{\rm Model~VII}:~{\cal L}_{\rm BSM}^{(Z)}=
-  \sum_{i,j} \left( {\lambda_i^{ *} \lambda_j m_Z^2 \over g_ZM^2}\right)
  \bigg( {\bar u}^{ i}_L \gamma^{\mu} u^{ j}_L 
+ {1 \over 2} \sum_{m,n} {\bar d}^{m}_L\, V^{\dagger}_{mi}\,
  \gamma^{\mu}\, V_{jn}\, d^{n}_L \bigg) Z_{\mu}\,.
\end{equation}
Similarly, in Model~VIII the result is
\begin{equation}
\label{newzcoup8}
{\rm Model~VIII}:~{\cal L}_{\rm BSM}^{(Z)}=
\sum_{i,j} \left( {\lambda_i^ * \lambda_j m_Z^2 \over g_Z M^2}\right)
  \bigg( {1\over 2}\, {\bar u}^{ i}_L \gamma^{\mu} u^{ j}_L
  + \sum_{m,n} {\bar d}^{m}_L\, V^{\dagger}_{mi}\,
  \gamma^{\mu}\, V_{jn}\, d^{ n}_L \bigg) Z_{\mu} \,.
\end{equation}
We can transform these $Z$ couplings to the basis where the down-type
quarks are mass eigenstates in the $\lambda_i \to 0$ limit by
redefining the couplings $\lambda_i \rightarrow \sum_k \lambda_k
V^{\dagger}_{ki}$. In this basis Eqs.~(\ref{newzcoup7}) and
(\ref{newzcoup8}) become
\begin{equation}
\label{newzcoup7a}
{\rm Model~VII} : {\cal L}_{\rm BSM}^{(Z)}= - \sum_{i,j} \left( 
  {\lambda_i^* \lambda_j m_Z^2 \over g_Z M^2}\right) \left(
 \sum_{m,n} {\bar u}^m_L\, V_{mi}\, \gamma^{\mu}\, V^{\dagger}_{jn} u^n_L 
  + \frac12\, \bar{d}^i_L\gamma^{\mu} d^j_L \right) Z_{\mu}\,,
\end{equation}
and
\begin{equation}
\label{newzcoup8a}
{\rm Model~VIII} : {\cal L}_{\rm BSM}^{(Z)} = - \sum_{i,j} \left( 
  {\lambda_i^* \lambda_j m_Z^2 \over g_Z M^2}\right) \left(
  \frac12 \sum_{m,n} { {\bar u}^m}_L\, V_{mi}\, \gamma^{\mu}\,
  V^\dagger_{jn} u^n_L 
  + \bar{d}^i_L\gamma^{\mu} d^j_L \right) Z_{\mu}\,.
\end{equation}

There are corrections to the $W$ boson couplings as well. We write the
couplings of the $W$-bosons to the left-handed light lepton and quark
mass eigenstates in terms of $3 \times 3$ matrices $X$ and $Y$
as\footnote{There are also right-handed currents in some of the
  models.},

\begin{equation}
{\cal L}^{(W)} = -{g_2 \over \sqrt{2}}\, W_{\mu}^+ \left[\, {\bar{ \nu}}{}_L^{i}\,
   \gamma^\mu X_{ij}\,
 { e}_L^{ j}  \right]  - {g_2 \over \sqrt{2}}\, W_{\mu}^+ \left[\, {\bar{u}}_L^{i}\,
 \gamma^\mu Y_{ij}\,
 { d}^ {j}_L  \right] +{\rm h.c.}\,.
\end{equation}
In the SM limit, $X$ is the inverse of the PMNS matrix and $Y$ is the
CKM matrix. Because of the mixing with the vector-like leptons in the
extensions of the SM discussed in this paper, $X$ and $Y$ are no longer
unitary matrices. It is straightforward to express $X$ and $Y$ in
terms of the components of the $4 \times 4$ diagonalization matrices
$V_{L,R}$ in the various models.

In this paper we focus on violations of lepton universality and
violations of unitarity of the CKM matrix. For that purpose we compute
the quantities.
\begin{align}
  &R_k \equiv\left( X X^{\dagger}\right)_{kk}=\sum_{j=1,2,3} X_{kj}X^{*}_{kj} \,, \\
  &S_{lm}\equiv\left(Y Y^{\dagger}\right)_{lm}= \sum_{j=1,2,3} Y_{lj}Y_{mj}^*\,,
\end{align}
at quadratic order in $(v^2/M^2)$, neglecting terms of higher
order. We find in the models with vector-like leptons that
\begin{align}
{\rm Model~I}:~ &R_k \simeq 1-| \lambda_k |^2  {v^2 \over M^2}\,,  \\
{\rm Model~II}:~&R_k \simeq 1-\left(
\frac{1}{2}|\lambda_k |^2+|\lambda_k^{\prime} |^2
-\lambda_k^*\lambda_k^{\prime}-\lambda_k\lambda_k^{\prime *}\right)
{v^2 \over M^2}  \,, \\
{\rm Models~III, IV}:~&R_k \simeq 1\,.
\end{align}
In Model II we have worked in a basis where the charged leptons ${\hat
  e}^j$ are mass eigenstates in the SM part of the charged lepton mass
matrix and so $\lambda^{\prime}_j=\sum_i \lambda_i U_{ij}$, where $U$
is the PMNS matrix.

In the models with vector-like quarks (neglecting the off diagonal
elements of the CKM matrix)
\begin{align}
{\rm Models~V,VI}:~&S_{lm} \simeq
\delta_{lm}-\lambda_l^{*} \lambda_m {v^2 \over M^2} \,, \\
{\rm Models~VII,VIII}:~&S_{lm} \simeq
\delta_{lm}+ {1 \over 2}\lambda_l^{*} \lambda_m {v^2 \over M^2}\,, \\
{\rm Models~IX, X, XI}:~&S_{lm} \simeq \delta_{lm}\,.
\end{align}
At this order in the $v^2/M^2$ expansion,
$T_{lm}\equiv(Y^{\dagger}Y)_{lm} =S_{lm}$.

\section{Experimental Constraints}

For leptonic-extension models, i.e., Models I -- IV, the constraints
from $\mu \rightarrow e$ conversion, $\mu \rightarrow 3e$ and $\tau
\rightarrow 3e$ are important. They are induced by the tree-level $Z$
couplings, which we derived in the previous section. On the other
hand, for hadronic-extension models, i.e., Models V -- XI, meson
mixing, such as $K^0$\,--\,$\bar{K}^0$ and $D^0$\,--\,$\bar{D}^0$, are
induced at one-loop level.  In addition, $\Delta F=1$ FCNC processes,
such as $K \rightarrow \pi \nu \bar{\nu}$, $K\to \mu^+\mu^-$,
$B_s\to\ell^+\ell^-$, etc., are induced at tree level. We will derive
effective Hamiltonians which are relevant for these processes and
discuss the current experimental bounds, as well as future prospects.
(For some recent studies of constraints on some of these models, see,
e.g., Refs.~\cite{Botella:2012ju, Fajfer:2013wca,
  Cacciapaglia:2015ixa}.)

For upper bounds on lepton flavor violating processes we quote bounds
at 90\% CL, as do most experiments, whereas for other measurements we
quote the $1\sigma$ limits that are approximately 84\% CL as one-sided
bounds for Gaussian distributions.

\subsection{Leptonic models}

The flavor violating tree-level couplings in
Eqs.~(\ref{newzcoup1})--\eqref{newzcoup4} give BSM tree-level
contributions to the flavor-changing neural-current processes; for example,
in the muon sector, to the $\mu \rightarrow 3e$ rate and muon conversion
to an electron in the presence of a nucleus. The amplitudes for these
processes are both proportional to the same combination of parameters
$\lambda_1 \lambda_{2 }^*/M^2$. The flavor diagonal terms in
Eqs.~(\ref{newzcoup1})--\eqref{newzcoup4} give rise to violations of
universality in lepton couplings to the $Z$. For typical couplings
$\lambda$ these are not as sensitive to large values of $M$ as the
flavor-changing charged lepton neutral currents are. Note that in
these models the radiative decay $\mu \rightarrow e \gamma$ does not
arise at tree level, only at one loop.

Let us first discuss $\mu \rightarrow e$ conversion in the presence of
a nucleus. It is among the most powerful probes of charged lepton
flavor violation beyond the standard model.  The $\mu^-$ conversion
rate to an $e^-$ in the presence of a nucleus $N$ is usually quoted as
a branching ratio normalized to the SM weak interaction, $\mu^- N
\rightarrow \nu_{\mu} N^{\prime}$ capture rate. At present the most
stringent bound is ${\rm Br}(\mu \rightarrow e{\rm ~conv.~in~Au}) < 7
\times 10^{-13}$ at the $90\%$ C.L.~\cite{mu2egold}. Future
experiments will have a dramatically improved sensitivity to a
branching ratio $ {\rm Br}( \mu \rightarrow e{\rm ~conv.~in~Al}) \sim
10^{-16}$~\cite{Abrams:2012er, Kurup:2011zza}.

The reach, in mass scale for new physics, of the next generation
charged lepton flavor violation (CLFV) experiments that search for
$\mu$ to $e$ conversion on Al and for the radiative decay $\mu
\rightarrow e \gamma$ was discussed in
Ref.~\cite{deGouvea:2013zba}. They assume an effective Lagrangian of
the form,
\begin{equation}
{\cal  L}_{\rm CLFV} =
{m_{\mu} \over (\kappa +1) \Lambda^2}\, {\bar \mu}_R \sigma_{\mu \nu}
F^{\mu \nu}e_L +
{\kappa  \over (\kappa +1) \Lambda^2}\, {\bar \mu}_L\gamma_{\mu}e_L
\left( {\bar u}_L \gamma^{\mu} u_L+ {\bar d}_L \gamma^{\mu} d_L \right)
+h.c.\,,
\end{equation}
and present the reach of these experiments in the $\kappa-\Lambda$
plane. For $\kappa \gg 1$ the limit on $\Lambda$ from conversion on
gold is close to $1 \times 10^3\,\TeV$ while future planned
experiments for conversion on aluminum are sensitive to $\Lambda
\simeq 7.2 \times 10^3\,\TeV$. In Models I -- IV the weak radiative
decay proceeds at the one-loop level and so indeed $\kappa \gg 1$.
Taking into account left-handed and right-handed quark currents which
couple to $Z$ boson, the relations between $\Lambda$ and the
vector-like lepton mass $M$ in Models $a=$ I, II, III, and IV are
\begin{equation}
\label{MvsLambda}
M^{{(a)}} =  \Lambda 
\left\{ \eta_{\mu\rightarrow e}^{(a)}\,
\frac{|\lambda_{2} \lambda_1|}3 \left[
  (2\sin^2\theta_W-1) \frac{Z}{A} + \frac12 \right] \right\}^{1/2} ,
\end{equation}
with $\eta_{\mu \rightarrow e}^{\rm (I)}=2\eta_{\mu \rightarrow
  e}^{\rm (II)}=\eta_{\mu \rightarrow e}^{\rm (III)}=\eta_{\mu
  \rightarrow e}^{\rm (IV)}=1$ and $Z$, $A$ are atomic number, mass
number of a nucleus, respectively. To obtain this expression we used
the vector part in the quark current, neglecting the axial
current. For Models III and IV, the right-handed lepton current occurs
but it gives the same contribution to the rate as the left-handed
one. On gold and aluminum Eq.~(\ref{MvsLambda}) implies that,
\begin{equation}\label{scaleconv}
M^{\rm (a)}_{\rm Au} = 0.31\Lambda\, 
   \sqrt{\eta_{\mu \to e}^{(a)}\,|\lambda_{2} \lambda_1|}\,, \qquad
   M^{\rm (a)}_{\rm A1} = 0.28 \Lambda\,
   \sqrt{ \eta_{\mu \to e}^{(a)}\, |\lambda_2 \lambda_1|}\,,
\end{equation}
We obtain from Fig.~2 of Ref.~\cite{deGouvea:2013zba} that a future
branching ratio limit ${\rm Br}(\mu \rightarrow e{\rm~conv.~in~Al}) <
10^{-16}$ will imply $\Lambda > 7.2\times 10^3\,\TeV$ for large
$\kappa$, yielding $M\big/ \sqrt{ \eta_{\mu \to e}^{(a)}\, |\lambda_2
  \lambda_1|} > 2.0\times 10^3\,\TeV$.  The current limit on $\mu \to
e$ conversion in Au implies that $\Lambda > 9.5\times
10^2\,\TeV$~\cite{deGouvea:2013zba}, which according to
Eq.~(\ref{scaleconv}) implies the limits,
\begin{equation} 
M^{(a)} > \sqrt{ \eta_{\mu \to e}^{(a)} \,|\lambda_2 \lambda_1|}
  \times 2.9\times 10^2\,\TeV\,.
\end{equation}

Charged lepton flavor violation is also constrained by the muon decay
$\mu \rightarrow 3e$ and similar $\tau$ decays. In the models $a=\{
{\rm I}, {\rm II}, {\rm III}, {\rm IV} \}$ the rates for these processes
normalized to a leptonic weak decay mode are,
\begin{equation}
{\Gamma(e^i \to e^j e^j {\bar e}^j) \over \Gamma(e^i \to \nu^i e {\bar \nu_e})}
={\eta_1^{(a)} \over 8G_F^2}\,
{\left| \lambda_i \lambda_j\right|^2 \over M^4}\,,
\end{equation}
and  
\begin{equation}
{\Gamma(e^i \to e^j e^k {\bar e}^k) \over \Gamma(e^i \to \nu^i e {\bar \nu_e})}
={\eta_2^{(a)} \over 8G_F^2}\,
{\left| \lambda_i \lambda_j\right|^2 \over M^4}\,.
\end{equation}
For the decays where the leptons in the final state are all of the same type,
\begin{equation}
\eta_1^{({\rm I})} = 4\eta_1^{({\rm II})} = 2\kappa_L^2+\kappa_R^2\,, \qquad
  \eta_1^{({\rm III})}=\kappa_L^2+2\kappa_R^2\,,
\end{equation}
where $\kappa_L= -1/2+\sin^2 \theta_W$ and $\kappa_R= \sin^2
\theta_W$, and $\theta_W$ is the Weinberg angle.  Numerically:
$\eta_1^{({\rm I})}\simeq0.20$, $\eta_1^{({\rm II})}\simeq 0.049$ and
$\eta_1^{({\rm III})}=\eta_1^{({\rm IV})} \simeq 0.18$. For the decays
where two types of leptons occur in the final state,
\begin{equation}
  \eta_2^{({\rm I})}= \eta_2^{({\rm III})}=\eta_2^{({\rm IV})}
  =\kappa_L^2+\kappa_R^2\,, \qquad
  \eta_2^{({\rm II})}={1 \over 4}\eta_2^{({\rm I})}\,.
\end{equation}
Numerically; $\eta_2^{({\rm I})}= \eta_2^{({\rm III})} =\eta_2^{({\rm
IV})}\simeq 0.13$ and $\eta_2^{({\rm II})} \simeq 0.031$. The 90\%
CL experimental limit, ${\rm Br}(\mu \rightarrow 3e)< 1.0 \times
10^{-12}$~\cite{Bellgardt:1987du} implies in Model~I that $M/\sqrt{ |
\lambda_1 \lambda_2 |}>1.2\times 10^2\, \TeV$.  While the limits
${\rm Br}(\tau \rightarrow 3 \mu) <2.1\times 10^{-8}$ and ${\rm
Br}(\tau \rightarrow 3 e) <2.7\times 10^{-8}$~\cite{Hayasaka:2010np}
imply in Model~I that $M/\sqrt{ | \lambda_2 \lambda_3 |} > 6.2\, \TeV$
and $M/\sqrt{ | \lambda_1 \lambda_3 |}>5.8 \, \TeV$.  The limits on
the rates for the $\tau$ decay channels $\tau \rightarrow \mu e {\bar
e}$ and $\tau \rightarrow e \mu {\bar \mu}$ give slightly weaker
limits because the $\eta_2$'s are smaller than the $\eta_1$'s.

The Mu3e experiment expects to reach in the absence of a signal the
90\% CL limit, ${\rm Br}(\mu \to 3 e) < 4\times
10^{-16}$~\cite{Bravar:2015vja}, yielding in Model~I, $M/\sqrt{ |
  \lambda_1 \lambda_2 |} > 8.2\times 10^2\, \TeV$.  The Belle~II
sensitivities for $\tau$ decays to three charged leptons are estimated
at the few times $10^{-10}$ level, and several channels will give
comparable sensitivity.  The bounds ${\rm Br}(\tau \to 3 e) < 4\times
10^{-10}$ and ${\rm Br}(\tau \to 3 \mu) < 4\times
10^{-10}$~\cite{Belle2predictions} would yield in Model~I, $M/\sqrt{ |
  \lambda_{1,2} \lambda_3 |} > 17\, \TeV$.

Competitive constraints to the above processes also arise from the upper bounds
on the $\tau \to e \pi$, $\tau \to \mu \pi$, $\tau \to e \rho$, and $\tau \to
\mu \rho$ branching ratios.  The vector-like fermions generate
\begin{equation}
\Gamma(\tau \to e^i\pi) = \eta_3^{(a)}\, \frac{|\lambda_3\lambda_i|^2}{M^4}\,
  \frac{m_\tau^3}{256\, \pi}\, f_\pi^2 \,,
\end{equation}
where $i=1,2$,
$\eta_3^{({\rm I})} = 4 \eta_3^{({\rm II})} = \eta_3^{({\rm III})}
  =\eta_3^{({\rm IV})} = 1$, that is,
$\eta_3^{(a)} = \eta_2^{(a)} / (\kappa_L^2+\kappa_R^2)$,
and we neglected $m_\pi^2/m_\tau^2$.
Similarly,
\begin{equation}
\Gamma(\tau \to e^i\rho) = \eta_3^{(a)}\, \frac{|\lambda_3\lambda_i|^2}{M^4}\,
  \frac{(m_\tau^2-m_\rho^2)^2\, (m_\tau^2+2m_\rho^2)}{256\, \pi\, m_\tau^3}\,
  \big(1-2\sin^2\theta_W\big)^2\, f_\rho^2 \,.
\end{equation}
Of the current 90\% CL experimental limits~\cite{Miyazaki:2008mw,
Hayasaka:2010et, Amhis:2014hma} the strongest  bounds arise from ${\rm Br}(\tau
\to e\pi)< 2.2 \times 10^{-8}$, which implies $M/\sqrt{ | \lambda_1 \lambda_3
|}>7.0\, \TeV$ in Model~I, and ${\rm Br}(\tau \to \mu\rho) < 1.2 \times
10^{-8}$, which implies $M/\sqrt{ | \lambda_2 \lambda_3 |} > 7.4\, \TeV$ in
Model~I.  The bounds from $\tau \to e\rho$ and $\tau \to \mu\pi$ are only
slightly weaker.

Concerning future sensitivity, the expected Belle~II limits 
are~\cite{Belle2predictions},
\begin{eqnarray}
{\rm Br}(\tau \to e \pi) &<& 4.0 \times 10^{-10}\,,\qquad
  {\rm Br}(\tau \to \mu \pi) < 4.9 \times 10^{-10}\,, \nonumber\\
{\rm Br}(\tau \to e \rho) &<& 3.1 \times 10^{-10}\,,\qquad
  {\rm Br}(\tau \to \mu \rho) < 2.0 \times 10^{-10}\,.
\end{eqnarray}
These imply that the strongest expected bounds in Model~I
will be $M/\sqrt{ | \lambda_1 \lambda_3 |}>19\, \TeV$ from $\tau \to e\pi$, and
$M/\sqrt{ | \lambda_2 \lambda_3 |} > 21\, \TeV$ from $\tau \to \mu\rho$.  The
expected sensitivities of the other channels are only slightly weaker, thus we
can have high confidence in the experimental reach, but not in which channel
will give the best bounds.

For generic BSM Yukawa couplings, $\lambda$, the reach for new physics
is much greater for experiments that search for charged lepton flavor
violation than those that seek violation of lepton universality.
However, it is possible for non-generic $\lambda$'s that the
violations of universality are more important. We close this
section by briefly commenting on the implications of the powerful
constraint on $e$-$\mu$ charged current universality coming from pion
decay $\pi^+\rightarrow e^+ \bar{\nu}(\gamma)$, $\mu^+
\bar{\nu}(\gamma)$. The latest experimental result gives
$g_e/g_{\mu}=0.9996 \pm 0.0012$~\cite{Aguilar-Arevalo} where
$g_{e,\mu}$ are the charged-current couplings of $e$ and $\mu$.\footnote{The
  constraint from Kaon decay~\cite{Ito:2015jwa, NA62:2011aa} is
  weaker.} In Models I -- IV, $g_e/g_{\mu}=\sqrt{R_1/R_2}$, which
implies, for example, in Model I that,
\begin{equation}
M >  \sqrt{\big| |\lambda_2|^2-|\lambda_1|^2\big|}\times 4.4~{\rm TeV} \,.
\end{equation}
It should be possible to improve this bound by a factor of two in the
future.  Tau decay, on the other hand, gives constraints on other
charged-current couplings, such as $g_\tau/g_{\mu}=1.0011 \pm
0.0015$~\cite{Amhis:2014hma}, which yields, e.g., in Model I,
\begin{equation}
M >  \sqrt{\big| |\lambda_2|^2-|\lambda_3|^2\big|}\times 6.2~{\rm TeV} \,.
\end{equation}

\subsection{Hadronic models}

To constrain Models V -- XI, FCNC processes in the quark sector are
most important.  (For a recent study of $Z$-mediated FCNC effects,
see, e.g., Ref.~\cite{Buras:2012jb}.)  We focus on leptonic and
semileptonic decays and neutral meson mixing.  In these models,
constraints from nonleptonic decays are weaker.

\subsubsection{Meson decays involving a $\nu\bar\nu$ pair}

For kaon decays involving a neutrino--antineutrino pair, such as
$K^+\to\pi^+\nu\bar\nu$, the effective Hamiltonian is
\begin{eqnarray}
\label{KpinunuSM}
{\cal H}& =& c_\nu \sum_i
  (\bar s_L \gamma_\mu d_L)(\bar\nu_{iL} \gamma^\mu \nu_{iL})
+ c'_\nu \sum_i (\bar s_R \gamma_\mu d_R)(\bar\nu_{iL} \gamma^\mu \nu_{iL})\, \\
&=& {c_{\nu,V} \over 2}\sum_i
(\bar s \gamma_\mu d)(\bar\nu_{iL} \gamma^\mu \nu_{iL}) +
{c_{{\nu},A}  \over 2}\sum_i
  (\bar s \gamma_\mu \gamma_5 d)(\bar\nu_{iL} \gamma^\mu \nu_{iL}) \,,
\end{eqnarray}
where in the second line of Eq.~(\ref{KpinunuSM}) the subscripts $V,A$
on the coefficients refer to the fact that these coefficients are for
the vector and axial quark currents and $ c_{\nu,V}= c'_\nu + c_\nu$,
$ c_{{\nu},A}= c'_\nu - c_\nu$. (See Appendix \ref{AppB}.)  Only the
vector part of the quark current contributes to this process. In the
SM $c_\nu'=0$ and at NNLO, $|c_{\nu,V}^{\rm (SM)}| =|c_\nu^{\rm (SM)}|
\simeq 1/(89\,\TeV)^2$, with a few percent
uncertainty~\cite{Buchalla:1993wq,Buchalla:1998ba,Buras:2004uu,Brod:2008ss,
  Buras:2015qea}. The hadronic models modify $c_{\nu,V}$
additively. Obviously in this case it is convenient to work in the
basis where the down-type quarks are mass eigenstates.\footnote{Going
  forward, for different processes we work in whatever basis makes the
  theoretical expressions simplest.  This means that the $\lambda$
  parameters are not always the same, but are linearly related through
  the CKM matrix.}  The tree-level $Z$ exchange BSM contributions to
the coefficient $c_{\nu,V}$ in models $a=$\,\{V, VII, VIII, IX, XI\}
have magnitude
\begin{equation}
|c_{\nu,V}^{(a)} |= \eta^{(a)}_{Zd}\,  {|\lambda_1 \lambda_2 |\over 2 M^2} \,,
\end{equation}
where for the FCNC down-type BSM contributions
\begin{equation}\label{etaZdown}
\eta_{Zd}^{({\rm V})} = 2\eta_{Zd}^{({\rm VII})} = \eta_{Zd}^{({\rm VIII})}
  = \eta_{Zd}^{({\rm IX})} = \eta_{Zd}^{({\rm XI})}=1.
\end{equation}
 Note that for Model~IX it is the $\lambda^{(d)}$'s that occur in this
 expression. Models VI and X do not contribute to this process through
 tree-level $Z$ exchange.

\begin{figure}[tb]
\includegraphics[width=.44\textwidth]{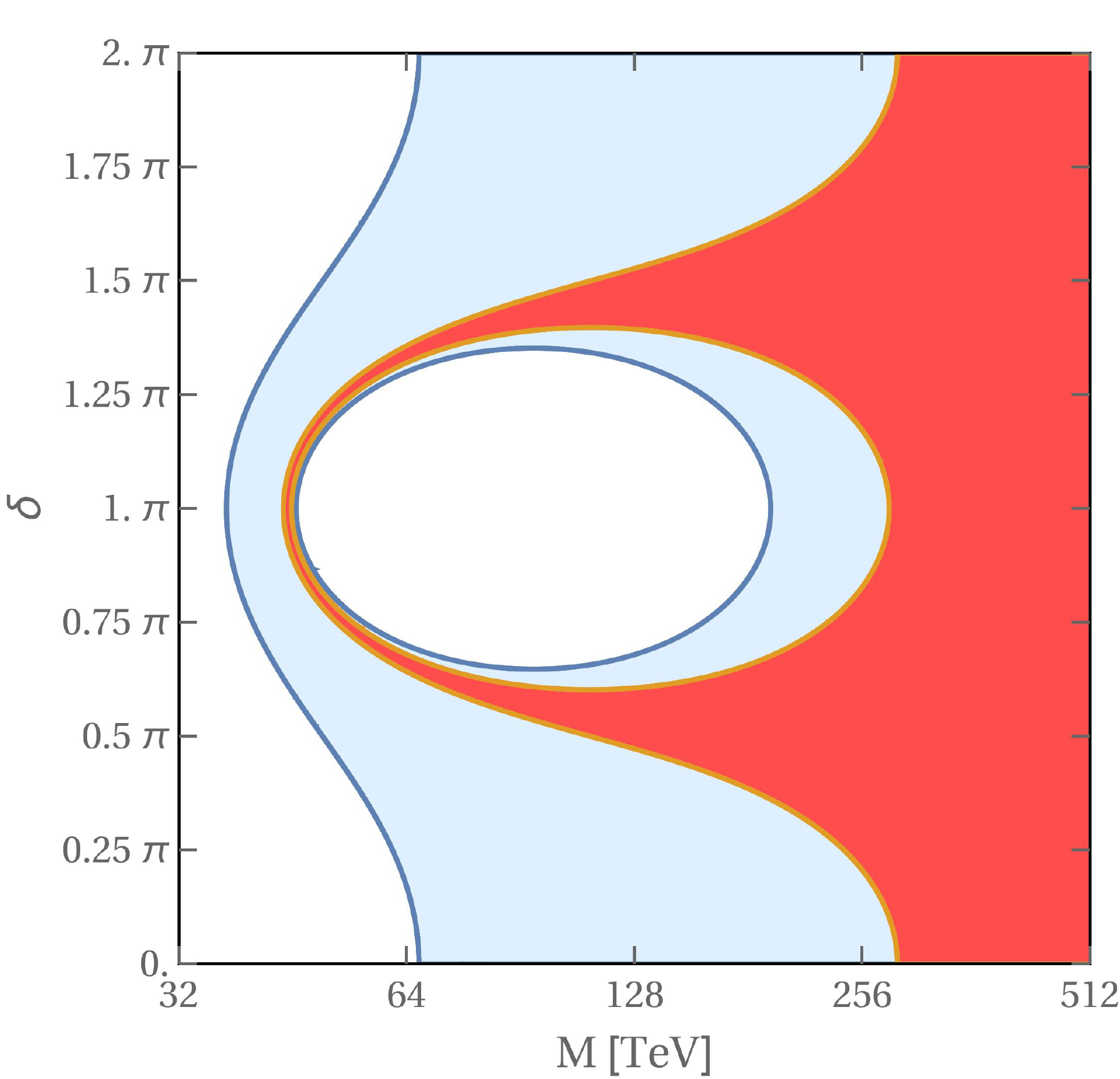}
\caption{Constraints from $K^+\to\pi^+\nu\bar\nu$ on vector-like
  fermions in the $M \big/ \sqrt{\eta_{Zd}^{({\rm a})}\, |\lambda_1
    \lambda_2|}$ vs.\ $\delta = \arg\big( c_{\nu,V}^{(a)} \big/
  c_{\nu,V}^{\rm (SM)}\big)$.  The currently allowed $1\sigma$ region is
  blue-shaded, corresponding to $0.8 < \big| 1 + c_{\nu,V}^{(a)} \big/
  c_{\nu,V}^{\rm (SM)} \big|^2 < 3.6$, whereas a future 10\%
  measurement in agreement with the SM, $0.9 < \big| 1 +
  c_{\nu,V}^{(a)} \big/ c_{\nu,V}^{\rm (SM)} \big|^2 < 1.1$, would
  constrain $M$ and $\delta$ to the red-shaded region.}
\label{fig:Kpinunu}
\end{figure}

The uncertainty of the SM prediction, ${\rm Br}^{\rm (SM)}(K^+ \to
\pi^+\nu\bar\nu) = (7.8 \pm 0.8) \times 10^{-11}$, is dominated by
that of $c_{\nu,V}^{\rm (SM)}$, and not by its relation to the
measured rate.  Hence, the measurement ${\rm Br}(K^+ \to
\pi^+\nu\bar\nu) = (1.7 \pm 1.1) \times
10^{-10}$~\cite{Artamonov:2008qb} implies $0.8 < \big| 1 +
c_{\nu,V}^{(a)} / c_{\nu,V}^{\rm (SM)} \big|^2 <
3.6$.\footnote{Despite the stated uncertainty, the probability that
  all 7 observed events were due to background was quoted as
  0.001~\cite{Artamonov:2008qb}.}  In Fig.~\ref{fig:Kpinunu} the
blue-shaded region shows the $1\sigma$ allowed region in the parameter
space of $M \big/ \sqrt{\eta^{(a)}_{Zd}\, {|\lambda_1 \lambda_2 |}}$
and $\delta = \arg\big( c_{\nu,V}^{(a)} \big/ c_{\nu,V}^{\rm
  (SM)}\big)$. If the BSM contribution is aligned (constructive
interference) with the SM, we find
\begin{equation}
\label{boundknunu}
M^{(a)} > \sqrt{\eta^{(a)}_{Zd}\, |\lambda_1 \lambda_2|} \times 66\,\TeV\,.
\end{equation}
This is not a bound in the strict sense.  Anti-alignment, $\delta \sim
\pi$, is possible for lower values of $M$. But it is more indicative
of the new physics reach, since a larger region in $\delta$ is allowed
for values of $M$ satisfying Eq.~(\ref{boundknunu}).

The upcoming 10\% measurement of ${\rm Br}(K^+ \to \pi^+\nu\bar\nu)$,
if consistent with the SM, would result in the ($1\sigma$) constraint
$0.9 < \big| 1 + c_{\nu,V}^{(a)} / c_{\nu,V}^{\rm (SM)} \big|^2 < 1.1$,
improving the sensitivity to BSM physics to about $2.8 \times
10^2\,\TeV$.  This is shown as the red-shaded region in
Fig.~\ref{fig:Kpinunu}.  Again we emphasize, as is clear from
Fig.~\ref{fig:Kpinunu}, that quoting the
weakest possible bound would not give the most useful impression of
the scale sensitivity, as a solution around $c_{\nu,V }^{(a)} = -2
c_{\nu,V}^{\rm (SM)}$ always remains, corresponding to a relatively low
new physics scale.  A $3\sigma$ deviation from the SM is possible for
$M^{(a)} \big/ \sqrt{\eta^{(a)}_{Zd}\, |\lambda_1 \lambda_2|} <
1.7\times 10^2\,\TeV$.

For the similar 3rd--2nd generation transition mediated by $b\to
s\nu\bar\nu$, the SM prediction for the coefficient of the operator
obtained from Eq.~(\ref{KpinunuSM}) by $d\to s$ and $\bar s\to \bar b$
replacements is $|c^{\rm (SM)}_{\nu ,V}| \simeq 1/(9.8\,\TeV)^2$
\cite{Buchalla:2000sk}.  This process has not been observed yet, and
the best current bound, ${\rm Br}(B\to K \nu\bar\nu) < 1.6\times
10^{-5}$, is about 4 times the SM
prediction~\cite{Altmannshofer:2009ma}.  This yields for new physics
aligned with the SM contribution, $M^{(a)} > \sqrt{\eta^{(a)}_{Zd}\,
  |\lambda_2 \lambda_3|} \times 6.9\,\TeV$.  In case the new physics
is anti-aligned, interfering destructively with the SM contribution,
$M^{(a)} > \sqrt{\eta^{(a)}_{Zd}\, |\lambda_2 \lambda_3|} \times
4.0\,\TeV$.  If the rate is at the SM level, Belle~II expects to
measure ${\rm Br}(B\to K^*\nu\bar\nu)$ with about 30\%
uncertainty~\cite{Belle2predictions},\footnote{At the SM level, $B\to
  K \nu\bar\nu$ is expected to get large backgrounds from $B\to K^*
  \nu\bar\nu$~\cite{Belle2predictions}.  While ${\rm Br}(B\to K
  \nu\bar\nu)$ depends only on $C_{+\nu}$, ${\rm Br}(B\to K^*
  \nu\bar\nu)$ also depends on $C_{-\nu}$, slightly complicating the
  analysis.} increasing the probed mass scales to about 19\,TeV (a
$3\sigma$ signal possible for $M^{(a)} \big/ \sqrt{\eta^{(a)}_{Zd}\,
  |\lambda_2 \lambda_3|} < $ $11\,\TeV$).

For $B\to \pi\nu\bar\nu$, Belle~II expects to reach a sensitivity at the
$1\times 10^{-5}$ level~\cite{Belle2predictions}, which will provide weaker
bounds than $B\to \pi\ell^+\ell^-$ in the models considered in this paper.

\subsubsection{Meson decays to an $\ell^+\ell^-$ pair}

The effective Hamiltonian differs from Eq.~(\ref{KpinunuSM}) in that
the coupling to the $\ell^+\ell^-$ pair can either be left- or
right-handed.  We write the effective Hamiltonian as
\begin{equation}
{\cal H} =\sum_{i =9,10} \left(c_i\, Q_i + c'_i\, Q'_i \right),
\end{equation}
where up to normalization the conventional choice of operator basis is
\begin{align}\label{Q910def}
Q_9 &= (\bar b_L \gamma_\mu s_L)(\bar\ell \gamma^\mu \ell)\,,
  \qquad\quad\  Q'_9 = (\bar b_R \gamma_\mu s_R)(\bar\ell \gamma^\mu \ell)\,,
  \nonumber\\
Q_{10} &= (\bar b_L \gamma_\mu s_L)(\bar\ell \gamma^\mu\gamma_5 \ell)\,,
  \qquad  Q'_{10} = (\bar b_R \gamma_\mu s_R)(\bar\ell \gamma^\mu\gamma_5
  \ell)\,,
\end{align}
with obvious replacements for $b\to d$ or $s\to d$ decays.  We use the
notation $c_i$ and $c'_i$ to emphasize that these are dimensionful
couplings, containing all terms multiplying the four-fermion operators
$Q_i$ in Eq.~(\ref{Q910def}).

\begin{figure}[tb]
\includegraphics[width=.44\textwidth]{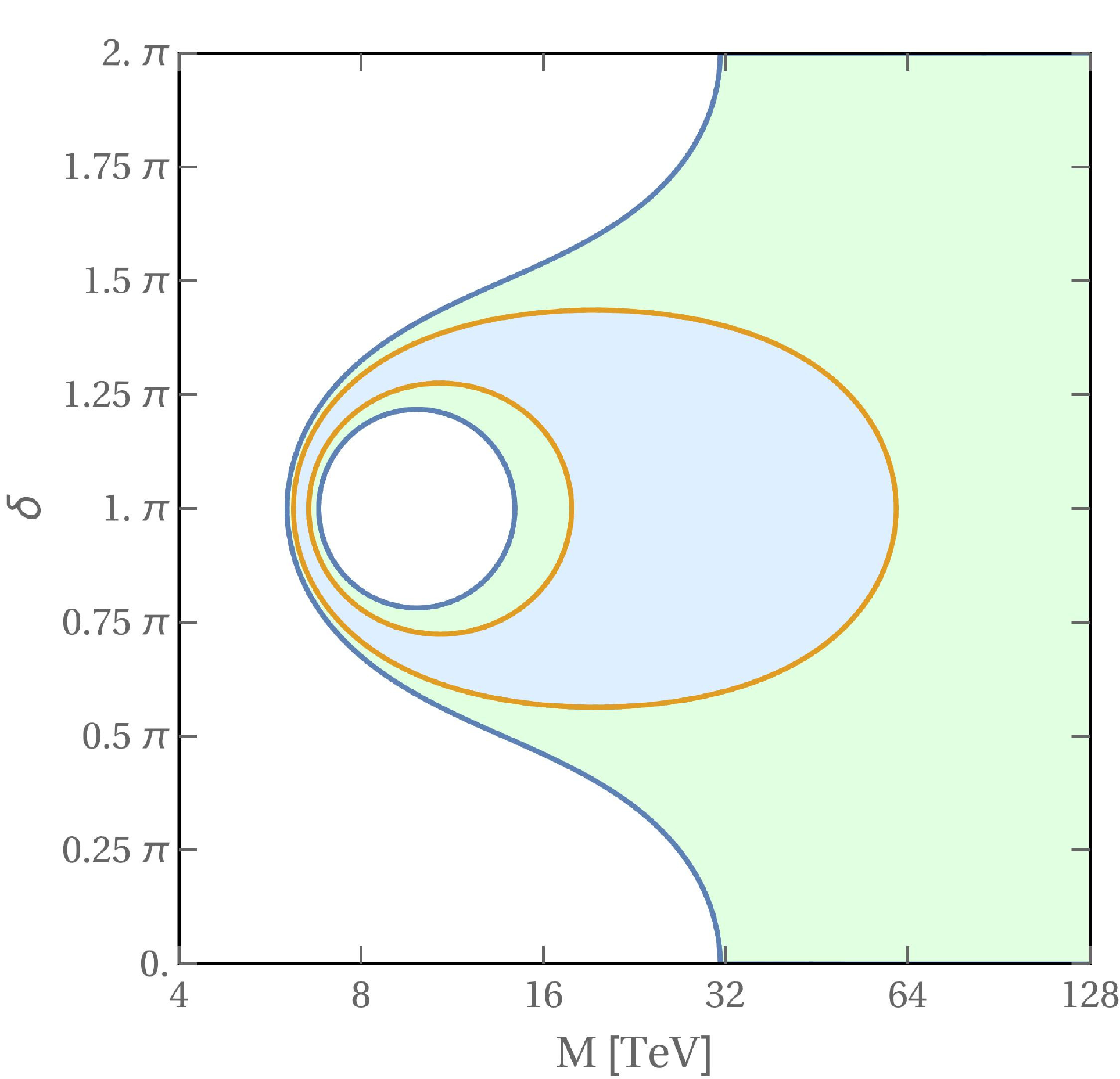}
\caption{Constraints from $B_s\to\mu^+\mu^-$ on vector-like fermions
  in the $M/\sqrt{ \eta_{Zd}^{({\rm a})}\, | \lambda_2 \lambda_3| }$
  vs.\ $\delta$ parameter space. The angle delta is defined in a
  similar way as in Fig.~\ref{fig:Kpinunu}.  The blue-shaded region is
  allowed at $1\sigma$ (indicating the mild tension with the SM), and
  the green-shaded region shows $2\sigma$.}
\label{fig:Bsmumu}
\end{figure}

For $B_s \to \mu^+ \mu^-$ the SM gives $|c_{10}^{({\rm SM})}| \simeq
1/(17\, \TeV)^2$ and $|c_{10}^{\prime ({\rm SM})} | \simeq 0$. The
amplitude for $B_s \rightarrow \mu^+ \mu^-$ is proportional to $
c_{10,A}=c_{10} - c'_{10}$. In the models with vector-like fermions
the BSM contributions to the coefficients $ c_{10,A}$ in models
$a=$\,\{V, VII, VIII, IX, XI\} have magnitude
\begin{equation}
|c_{10, A}^{(a)}| = \eta^{(a)}_{Zd}\, {|\lambda_2 \lambda_3| \over 4 M^2}\,,
\end{equation}
and the $\eta_{Zd}$ coefficients are the same as in
Eq.~(\ref{etaZdown}).  The LHCb--CMS combination of their
measurements, ${\rm Br}(B_s\to \mu^+\mu^-) = (2.8^{+0.7}_{-0.6})
\times 10^{-9}$, is quoted as the SM prediction times
$0.76^{+0.20}_{-0.18}$~\cite{CMS:2014xfa}. The $1\sigma$ range of the
measured $B_s\to \mu^+\mu^-$ rate is slightly outside the SM and
corresponds to the region, $0.96>|1+c_{10, A}^{(a)}/c_{10, A}^{({\rm
SM})}|^2>0.58$. The allowed one and two sigma regions of BSM
parameter space are plotted in Fig.~\ref{fig:Bsmumu} using variables
very similar to Fig.~\ref{fig:Kpinunu}.
If the BSM contribution is destructive (at $1\sigma$),
\begin{equation}\label{BsmumuBound}
M^{(a)} > \sqrt{\eta^{(a)}_{Zd}\, |\lambda_2 \lambda_3|} \times 18\, \TeV\,.
\end{equation}
A future 10\% measurement~\cite{LHCb-PUB-2014-040, CMS_Bmumu_future}
would increase this sensitivity to $\sim$\,40\,TeV. Again we note that
Eq.~(\ref{BsmumuBound}) is not a true bound on $M$ but rather is meant
to give a feeling for the reach in $M$ of current experimental data on
this decay mode. There is a tuned region of BSM parameter space near
$c_{10, A}^{(a)}/c_{10, A}^{({\rm SM})}=-2$ that corresponds to a
lower value of $M$.

For $B_d\to\mu^+\mu^-$ the SM prediction is $|c_{10,A}^{({\rm SM})}|
\simeq 1/(37\, \TeV)^2$.  The LHCb--CMS combination, ${\rm Br}(B_d\to
\mu^+\mu^-) = (3.9^{+1.6}_{-1.4}) \times 10^{-10}$ is quoted as the SM
prediction times $3.7^{+1.6}_{-1.4}$~\cite{CMS:2014xfa}.  If the new
contribution is constructive to the SM, then
\begin{equation}\label{BdmumuBound}
M^{(a)} > \sqrt{\eta^{(a)}_{Zd}\, |\lambda_1 \lambda_3|} \times 16\, \TeV\,.
\end{equation}
Accommodating the current central value would require $M^{(a)} /
\sqrt{\eta^{(a)}_{Zd}\, |\lambda_3\lambda_1|} \simeq 19\,\TeV$.  In
the HL-LHC era, this measurement will reach an uncertainty around 20\%
of the SM prediction~\cite{LHCb-PUB-2014-040, CMS_Bmumu_future},
increasing the mass reach to above $60$\,TeV.

The extraction of the short-distance part of the measured rate ${\rm
  Br}(K_L\to\mu^+\mu^-) = (6.84 \pm 0.11)\times 10^{-9}$~\cite{PDG} is
subject to considerable uncertainties.  The estimate ${\rm
  Br}(K_L\to\mu^+\mu^-)_{\rm SD} \leq 2.5\times
10^{-9}$~\cite{Isidori:2003ts} is about 3 times the SM short-distance
rate which follows from $|c_{10,A}^{\rm SM}|=1/(1.8\times 10^2\,
\TeV)^2$. If the BSM component is aligned with the SM, we find
\begin{equation}\label{KLmumuBound}
M^{(a)} > \sqrt{\eta^{(a)}_{Zd}\,| {\rm Re}(\lambda_1
  \lambda_2^*)}| \times 100\,\TeV \,.
\end{equation}
Unlike the bound from $K^+\to \pi^+\nu\bar\nu$, the prospect of
improving this is not good, and the uncertainties are greater.

Recently LHCb established a strong bound ${\rm Br}(D^0\to \mu^+\mu^-) < 6.2
\times 10^{-9}$ at 90\% CL~\cite{Aaij:2013cza}, which is well above
the SM level.  In Models~VI--X,
\begin{equation}
{\rm Br}(D^0\to \mu^+\mu^-) = \tau_{D^0}\, 
  \frac{|\lambda_1\lambda_2|^2}{M^4}\, \frac{[\eta_{Zu}^{(a)}]^2}{128\,\pi}\, 
  f_D^2\, m_D\, m_\mu^2 \sqrt{1-4m_\mu^2/m_D^2} \,,
\end{equation}
where
\begin{equation}\label{etaZup}
\eta_{Zu}^{\rm (VI)} = \eta_{Zu}^{\rm (VII)}
  = 2\, \eta_{Zu}^{\rm (VIII)} = \eta_{Zu}^{\rm (IX)} 
  = \eta_{Zu}^{\rm (X)} = 1 \,.
\end{equation}
 For Models~VII and VIII we used the basis in Eqs.~(\ref{newzcoup7}) and
 (\ref{newzcoup8}).  Using $f_D = 0.209\,\GeV$~\cite{Aoki:2013ldr}, we
 obtain
\begin{equation}
M^{(a)} > \sqrt{\eta^{(a)}_{Zu}\, |\lambda_1 \lambda_2|} \times 3.9\, \TeV\,.
\end{equation}
It is possible that in the HL-LHC era the experimental bound will
improve by a factor of $\sim 20$~\cite{Tim}.

\subsubsection{Semileptonic decays to $\ell^+\ell^-$ pairs}

The bounds on $K_L\to \pi^0\ell^+\ell^-$ are about an order of
magnitude above the SM expectation, so the resulting constraints are
weaker than those obtained from $K\to\pi\nu\bar\nu$ and
$K_L\to\mu^+\mu^-$.

LHCb recently measured ${\rm Br}(B^+\to \pi^+\mu^+\mu^-) = (2.3 \pm
0.6) \times 10^{-8}$~\cite{LHCb:2012de}, consistent with the SM
prediction quoted as $(2.0 \pm 0.2) \times 10^{-8}$.  This was the
first FCNC $b\to d$ decay observed (other than $B_d$ mixing).
Requiring that the SM rate is not enhanced by more than 50\% by new
physics, using Appendix~\ref{AppB}, we find
\begin{equation}
M^{(a)} > \sqrt{\eta^{(a)}_{Zd}\, |\lambda_1 \lambda_3|} \times 30\, \TeV\,.
\end{equation}
This bound is slightly stronger that that from $B_d \to \mu^+\mu^-$,
moreover, the current $B^+\to \pi^+\mu^+\mu^-$ measurement only used
0.9/fb data.  As the measurement gets more precise, a dedicated
analysis of $B^+\to \pi^+\mu^+\mu^-$, possibly considering $[{\rm
    d}\Gamma(B^+\to \pi^+\mu^+\mu^-)/{\rm d}q^2] / [{\rm d}\Gamma(B\to
  \pi \ell\bar\nu)/{\rm d}q^2]$ to reduce theoretical uncertainties,
is warranted.

The analysis of $B\to K\mu^+\mu^-$ is very similar to $B^+\to
\pi^+\mu^+\mu^-$, while $B\to K^*\mu^+\mu^-$ is more complicated and
does not give better bounds.  Averaged over $\ell=e,\mu$, HFAG quotes
${\rm Br}(B\to K\ell^+\ell^-) = (4.8 \pm 0.4) \times
10^{-7}$~\cite{Amhis:2014hma}.  The experimental uncertainty is
smaller than the theoretical one (due to the form factors).  Demanding
a less than 30\% modification of the SM rate, we find $M^{(a)} >
\sqrt{\eta^{(a)}_{Zd}\, |\lambda_2 \lambda_3|} \times 17\, \TeV$.
This happens to be very close to the bound in Eq.~(\ref{BsmumuBound}),
but the prospects of improving that are better.

The inclusive decay rate ${\rm Br}(B\to X_s\ell^+\ell^-) = (5.0 \pm
0.6) \times 10^{-6}$~\cite{Amhis:2014hma} depends in the models we
consider on $|c_9|^2 + |c'_9|^2$ and $|c_{10}|^2 + |c'_{10}|^2$.  In
Models~IX and XI, there is no interference between the SM and the new
physics contributions, so the constraints are weak.  We follow
Ref.~\cite{Huber:2015sra}, which studied the rates in the low- and
high-$q^2$ regions and found that for the SM value of $c_9$ the
constraint on $c_{10}$ in Models~V, VII, and VIII is $0.88 < 1 +
c_{10}^{(a)}/c_{10}^{\rm (SM)} < 1.17$ (at $1\sigma$).  Here it is
assumed that $c_{10}^{(a)}/c_{10}^{\rm (SM)}$ is real. This implies
\begin{equation}
M^{(a)} > \sqrt{\eta^{(a)}_{Zd}\, |\lambda_2 \lambda_3|} \times 21\, \TeV\,.
\end{equation}

In the up-quark sector, LHCb established a bound ${\rm Br}(D^+\to
\pi^+\mu^+\mu^-) < 7.3 \times 10^{-8}$ at 90\% CL~\cite{Aaij:2013sua},
removing regions of $m_{\mu^+\mu^-}$ with resonance contributions. We
find that this bound is weaker on the models considered in this paper
than the one from ${\rm Br}(D^0\to \mu^+\mu^-)$, and the latter is
also theoretically cleaner.

For FCNC top decays, CMS has set the best bound so far, ${\rm Br}(t\to
qZ) < 5\times 10^{-4}$~\cite{Chatrchyan:2013nwa} at 95\% CL (where
$q=c,u$, corresponding to $i=2,1$ below, respectively).  Comparing to
the dominant $t\to bW$ rate, in models VI--X,
\begin{equation}
{\Gamma(t\to q Z) \over \Gamma(t\to bW)} \simeq 
  {2| \eta_{Zu}^{(a)}\,\lambda_3 \lambda_i|^2 \over g_Z^4}\,
  {m_Z^4\over M^4}\, .
\end{equation}
For Models~VII and VIII we used again the basis in
Eqs.~(\ref{newzcoup7}) and (\ref{newzcoup8}).  We find
$M^{(a)}/\sqrt{\eta_{Zu}^{(a)}\, |\lambda_3\lambda_i|} > 0.96\,\TeV$.
The HL-LHC is expected to reach sensitivity at the $10^{-5}$
level~\cite{ATLAS:2013hta}, which will improve this bound to about
2.3\,\TeV.  However, the direct (and $\lambda$-independent) searches
are comparably sensitive, and are expected to remain to be so.

\subsubsection{Neutral meson mixing}
\label{sec:mix}

\begin{figure}[tb]
\includegraphics[width=.3\textwidth]{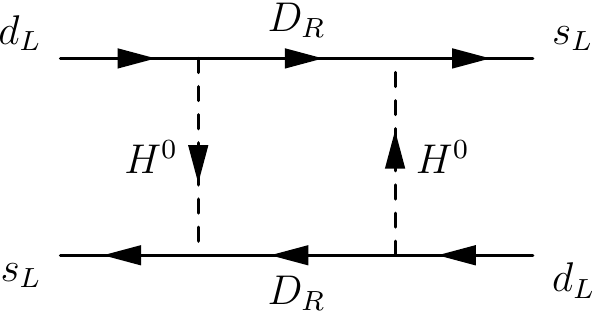} \hfil
\includegraphics[width=.3\textwidth]{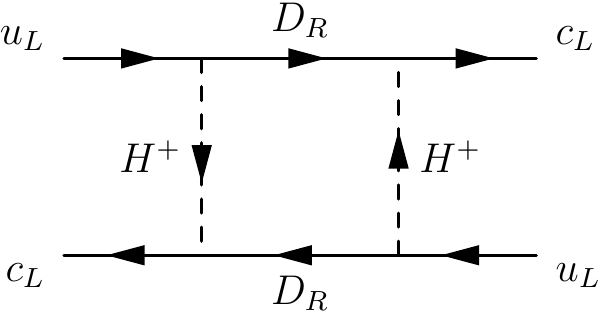}
\caption{Diagrams contributing to $K$ and $D$ mixing in Model~V.}
\label{fig:mixing}
\end{figure}

Since the new fermions interact with the Higgs field and the ordinary
light quarks via Yukawa couplings, meson mixing, such as
$K^0$\,--\,$\bar{K}^0$, $D^0$\,--\,$\bar{D}^0$, or
$B^0$\,--\,$\bar{B}^0$, is induced. The effective Hamiltonian for
these processes contains dimension-six four-quark operators with
coefficients of mass dimension $-2$.  At tree level, through $Z$
exchange, the coefficients are of the form $\sim ( \lambda_i
\lambda_j^{*})^2 v^2/M^4$.  However, coefficients of order $(
\lambda_i \lambda_j^{*})^2 /(4 \pi M)^2$ are generated at one loop
that are not CKM and/or quark-mass suppressed.  For large $M$, these
one-loop matching contributions are more important than tree-level $Z$
exchange.  Furthermore, they are independent of the Higgs vacuum
expectation value, $v$, and arise from short distances $\sim
1/M$. They can be calculated in the symmetric phase and come from box
diagrams with virtual scalars and the heavy vector-like fermions in
the loop; see Fig.~\ref{fig:mixing}.  The resulting effective
Lagrangians are,
\begin{align}
&{\cal L}_{\rm meson}^{(a)}= - \eta^{(a)}_{\rm mix}\, 
  {(\lambda_i^* \lambda_j )^2\over 128 \pi^2 M^2}
  \bigg[ \sum_{klmn} \big( {\bar u}_L^k V_{ki}\, \gamma_{\mu} V^{\dagger}_{jl}\,
  u_L^l \big) 
\big( {\bar u}_L^m V_{mi}\,\gamma_{\mu} V^{\dagger}_{jn}u_L^n \big)
  + \big( {\bar d}_L^i \gamma_{\mu}d_L^j \big)
  \big( {\bar d}_L^i \gamma^{\mu}d_L^j \big) \bigg] + h.c. ,\\
&{\cal L}_{\rm meson}^{\rm (IX)} = -
  {\big(\lambda^{(u)}_i \lambda_j^{(u)*}\big)^2 \over 64 \pi^2 M^2}
  \big( {\bar u}_L^i \gamma_{\mu}u_L^j \big)
  \big( {\bar u}_L^i \gamma^{\mu}u_L^j \big)
+{\big(\lambda^{(d)}_i \lambda_j^{(d)*}\big)^2 \over 64 \pi^2 M^2}
  \left( {\bar d}_L^i \gamma_{\mu}d_L^j \right)
  \left( {\bar d}_L^i \gamma^{\mu}d_L^j \right)+h.c. ,\\
&{\cal L}_{\rm meson}^{\rm (X)}= -
  {(\lambda_i \lambda_j^*)^2 \over 128 \pi^2 M^2}
  \left( {\bar u}_R^i \gamma_{\mu}u_R^j \right)
  \left( {\bar u}_R^i \gamma^{\mu}u_R^j \right)+h.c. ,\\
&{\cal L}_{\rm meson}^{\rm (XI)}= -
  {(\lambda_i \lambda_j^*)^2 \over 128 \pi^2 M^2}
  \left( {\bar d}_R^i \gamma_{\mu}d_R^j \right)
  \left( {\bar d}_R^i \gamma^{\mu}d_R^j \right)+h.c. ,
\end{align}
with $\eta_{\rm mix}^{\rm (V)}=\eta_{\rm mix}^{\rm (VI)}=1$ and
$\eta_{\rm mix}^{\rm (VII)}=\eta_{\rm mix}^{\rm (VIII)}=5/4$.  In the
Lagrangians for Models~V -- VIII the down-type quarks are the SM mass
eigenstates. For the remaining models the quark fields occurring both
in their Lagrangians and in the effective Lagrangians for meson mixing
can be taken to be SM mass eigenstates.

The operators above are renormalized with the subtraction point at the
scale $M$. They can be related through a QCD correction factor $\xi$
to scale invariant operators whose matrix elements can be evaluated
using lattice QCD.  So we write
\begin{equation}
\langle {\bar P }| ({\bar q}_{(L,R)} \gamma^{\mu}q_{(L,R)})^2
  |P \rangle = {2 \over 3}\, \xi {\hat B_{P}} f_P^2\, m_P^2 \,,
\end{equation}
where $\hat B_P$ does not depend on subtraction point.  For $K$
mixing, $f_K=156\,\MeV$ and ${\hat B_K}\simeq 0.76$.  In that case the
QCD correction factor, $\xi$, is given in the leading logarithmic
approximation~\cite{Gilman:1980di} by
\begin{equation}
\xi = \left[{\alpha_s(M)\over \alpha_s(m_t)} \right]^{6/21}
  \left[ { \alpha_s(m_t) \over \alpha_s(m_b)} \right]^{6/23}
  \left[ { \alpha_s(m_b) \over \alpha_s(\mu_0)} \right]^{6/25}
  [\alpha_s(\mu_0)]^{6/25} \simeq [\alpha_s(M)]^{6/21}.
\end{equation}
The scale $\mu_0$ does not affect any physical results. For $10\,\TeV
< M < 100\,\TeV$, $\xi \simeq 0.5$, and this value is approximately
the same for mixing in the $K$, $D$, and $B_{d,s}$ systems.  We use
the lattice QCD averages from Ref.~\cite{Aoki:2013ldr}, and $\hat B_D
= B_D^{\overline{\rm MS}}(3\,\GeV) / [\alpha_s(3\,\GeV)]^{6/25} \simeq
1.0$ from a recent calculation of $B_D^{\overline{\rm
    MS}}(3\,\GeV)$~\cite{Carrasco:2015pra}.

In the case of kaon mixing, we demand the new physics contribution to
the real part of $M_{12}$ to be less than 40\% of the experimental
value of $\Delta m_K$~\cite{Bai:2014cva}.  We take the $1\sigma$
uncertainty on the determination of $\epsilon_K$ in the SM to be
$25\%$ (reading off the bound $h_K<0.35$ from
Ref.~\cite{Charles:2013aka}).  For all the (hadronic) models, except
Model~X, this results in the constraint
\begin{equation}\label{mixboundK}
M^{(a)} > \max \left[ 42\, \sqrt {\eta_{\rm mix}^{(a)}\, 
  \big|{\rm Re}(\lambda_2\lambda_1^*)^2\big|} \,,\
  6.7 \times 10^2\, \sqrt{ \eta_{\rm mix}^{(a)}\, 
  \big|{\rm Im}(\lambda_2\lambda_1^*)^2\big|} \right] \TeV\,.
\end{equation}
(Note that the 90\% CL constraint from $\epsilon_K$ is only slightly
weaker, replacing $6.7 \times 10^2\,\TeV$ by $6.3 \times 10^2\,\TeV$.)
In the future, the $\epsilon_K$ constraint is expected to improve by
about $\sqrt{2.4}$~\cite{Charles:2013aka}, replacing $6.7 \times
10^2\,\TeV$ by $1.0 \times 10^3\, \TeV$ in Eq.~(\ref{mixboundK}).
Improvement in the real part is contingent upon lattice QCD
calculations of the long-distance contributions to $\Delta m_K$;
reaching $x\%$ precision would replace 42\,TeV in
Eq.~(\ref{mixboundK}) by $260\,\TeV/\sqrt{x}$.

In $B_{d,s}$ meson mixing, the new physics contribution is
conventionally parametrized as $M_{12}^{d,s} = (M_{12}^{d,s})_{\rm SM}
\times \big(1 + h_{d,s}\, e^{2i\sigma_{d,s}}\big)$.  Until recently,
the bounds on real ($2\sigma = 0$, mod\,$\pi$, that is MFV-like) and
imaginary ($2\sigma = \pi/2$, mod\,$\pi$) new physics contributions
have been quite different~\cite{Blum:2009sk, Isidori:2010kg}.  This is
no longer the case~\cite{Charles:2013aka}, and since we are most
interested in physics reach, we simply quote the limits on the
absolute values of the new physics contribution.  In $B_{d,s}$ mixing,
$h_d<0.3$ and $h_s<0.2$~\cite{Charles:2013aka} yield in all models
except Model~X, using $f_{B_d}=188\,\MeV$, $f_{B_s}=226\,\MeV$,
$\hat{B}_{B_d}=1.27$, $\hat{B}_{B_s}=1.33$~\cite{Aoki:2013ldr},
\begin{equation}\label{mixboundB}
M^{(a)} > \sqrt{\eta_{\rm mix}^{(a)}\, 
  |\lambda_3\lambda_1|^2} \times 25\, \TeV\,, \qquad
M^{(a)} > \sqrt{\eta_{\rm mix}^{(a)}\, 
  |\lambda_3\lambda_2|^2} \times 6.4\, \TeV\,.
\end{equation}
(For the imaginary part in $B_d$ mixing, 25\,TeV should be replaced by
31\,TeV, which is a smaller dependence on the phase of new physics
than those ignored for $\Delta F=1$ FCNC transitions earlier.)  In the
above constraints $\eta_{\rm mix}^{({\rm IX})} =2$, $\eta_{\rm
mix}^{({\rm XI})} =1$, and for Model~IX it is the $\lambda^{(d)}$'s
that occur in the constraints.  In the next decade these limits will
improve to $h_d < 0.05$ and $h_s <
0.04$~\cite{Charles:2013aka}, which will replace 25\,TeV by 61\,TeV
and 6.4\,TeV by 14\,TeV in Eq.~(\ref{mixboundB}).

The mixing of $D$ mesons is probably dominated by long-distance
physics.  Thus, we can only require that the new physics contribution
does not exceed the measurement, i.e., $\Delta m_D/\Gamma_D <
0.006$~\cite{Amhis:2014hma}.  (The significance of $\Delta m_D\neq 0$
is less than $2\sigma$, so we use the upper bound of the $1\sigma$
region.  We do not distinguish between imaginary and real
contributions to $M_{12}$ relative to the SM; the bounds on the
mass scale may differ by a factor $\sim 2$, depending on the value of 
$\Delta m_D$.)  In contrast to above, the $\lambda^{(u)}$ couplings
occur in the constraint for Model~IX.  Furthermore, in this case we
choose the up-type quarks to be the SM mass eigenstate fields, so the
factors of the CKM matrix move to the terms with the down-type
quarks. Then $D$ mixing implies for all models except Model~XI,
\begin{equation}\label{mixboundD}
M^{(a)} > \sqrt {\eta_{\rm mix}^{(a)}\, |\lambda_2\lambda_1|^2} \times
  48\, \TeV\,,
\end{equation}
where now $\eta_{\rm mix}^{({\rm X})}=1$. Note that for Models~V--VIII
the constraint from $D$ mixing is slightly stronger than from the real
part of $K$ mixing. However, the constraints are actually a little
different since the $\lambda$'s in Eqs.~(\ref{mixboundB}) and in
(\ref{mixboundD}) are not the same.  They are linearly related through
the CKM matrix.  The future evolution of this bound is uncertain.  While 
Belle~II expects to measure $\Delta m_D/\Gamma_D$ with an uncertainty
of 0.001~\cite{Belle2predictions}, the central value will matter for
the bounds on new physics, and therefore we do not assume that this bound 
will improve.

\subsubsection{Unitarity of the CKM Matrix}

The CKM matrix is not unitary in the extensions of the SM we are
considering. The strongest constraints on violations of unitarity come
form the first row and first column of the CKM matrix~\cite{PDG}
\begin{equation} 
 S_{11} =0.9999 \pm 0.0006 \quad {\rm and} \quad T_{11}=1.000 \pm0.004\,.
\end{equation}
In the models with vector-like quarks both of these constraints
involve the same combination of couplings and we find
\begin{equation}
M^{(a)} > |\lambda_1| \sqrt{\eta_{\rm un}^{(a)}}\times 7.8~{\rm TeV}\,,
\end{equation}
where in $\eta_{\rm un}^{\rm (V)}=\eta_{\rm un}^{\rm (VI)}=2\eta_{\rm
  un}^{\rm (VII)}=2\eta_{\rm un}^{\rm (VIII)}=1$ and $\eta_{\rm
  un}^{\rm (IX)}=\eta_{\rm un}^{\rm (X)}=\eta_{\rm un}^{\rm
  (XI)}=0$. For the second row we have the constraint
\begin{equation}
 S_{22} =1.002 \pm 0.027\,, 
\end{equation}
which yields
\begin{equation}
M^{(a)} > |\lambda_2| \sqrt{\eta_{\rm un}^{(a)}}\times 1.1~{\rm TeV}\,.
\end{equation}.

\section{Conclusions}

There are 11  renormalizable models that add to the SM
vector-like fermions in a single (complex) representation of the gauge
group that can Yukawa couple to the SM fermions through the Higgs
field.   These BSM fermions can have a mass
$M$ that is much greater than the weak scale, since they have a mass
term even in the absence of weak symmetry breaking. However, unlike
BSM scalars, such fermions are technically natural.  These models are
a class of very simple extensions of the SM that do not worsen the SM
hierarchy puzzle.

The masses of these vector-like fermions can take any value, up to the
ultraviolet cutoff ($\sim M_{\rm Pl}$), and so there is no particular
reason that they should be in a region that can be probed
experimentally. However, there are many such models and it is not
unreasonable that one of them has vector-like fermions with masses in
the experimentally testable range.

\begingroup
\begin{table*}[tb] \tabcolsep 6pt
\begin{tabular}{c@{\extracolsep{2pt}}l|cc|cc|cc}
\hline\hline
\multirow{2}{*}{Model} &  \multicolumn{1}{c|}{Quantum}  &
  \multicolumn{6}{c}{Present bounds on $M/\TeV$ and $\lambda_i \lambda_j$
  for each $ij$ pair}  \\
& \multicolumn{1}{c|}{numbers} & \multicolumn{2}{c|}{$ij=12$} 
  & \multicolumn{2}{c|}{$ij=13$} & \multicolumn{2}{c}{$ij=23$}  \\
\hline
I & $(1,1,-1)$ & \multicolumn{2}{c|}{310$^a$} & \multicolumn{2}{c|}{7.0$^b$}
 & \multicolumn{2}{c}{7.4$^c$}
\\
II & $(1,3,-1)$ & \multicolumn{2}{c|}{220$^a$} & \multicolumn{2}{c|}{4.9$^b$}
 & \multicolumn{2}{c}{5.2$^c$}
\\
III & $(1,2,-1/2)$ & \multicolumn{2}{c|}{310$^a$} & \multicolumn{2}{c|}{7.0$^b$}
  & \multicolumn{2}{c}{7.4$^c$}
\\
IV & $(1,2,-3/2)$ & \multicolumn{2}{c|}{310$^a$} & \multicolumn{2}{c|}{7.0$^b$}
  & \multicolumn{2}{c}{7.4$^c$}
\\ \cline{3-8}
& & $\Delta F=1$ & $\Delta F=2$ & $\Delta F=1$ & $\Delta F=2$
  & $\Delta F=1$ & $\Delta F=2$
\\ \cline{3-8}
V & $(3,1,-1/3)$ & 66$^d$ [100]$^e$ & \{42, 670\}$^f$
  & 30$^g$ & 25$^h$ & 21$^i$ & 6.4$^j$
\\
VI & $(3,1,2/3)$ & 3.9$^k$ & \{42, 670\}$^f$ & --- & 25$^h$ & --- & 6.4$^j$
\\
VII & $(3,3,-1/3)$  & 47$^d$ [71]$^e$ & \{47, 750\}$^f$ 
  & 21$^g$ & 28$^h$ & 15$^i$ & 7.2$^j$
\\
VIII & $(3,3,2/3)$  & 66$^d$ [100]$^e$ & \{47, 750\}$^f$
  & 30$^g$ & 28$^h$ & 21$^i$ & 7.2$^j$
\\
IX $\lambda^{(u)}$ & \multirow{2}{*}{$\bigg\}(3,2,1/6)$}
  & 3.9$^k$ & 67$^l$ & --- & 35$^h$ & --- & 9.1$^j$
\\
IX $\lambda^{(d)}$ &  & 66$^d$ [100]$^e$ & \{59, 950\}$^f$
  & 30$^g$ & 35$^h$ & 18$^m$ & 9.1$^j$
\\
X & $(3,2,7/6)$ & 3.9$^k$ & 48$^l$ & --- & --- & --- & ---
\\
XI & $(3,2-5/6)$  & 66$^d$ [100]$^e$ & \{42, 670\}$^f$
  & 30$^g$ & 25$^h$ & 18$^m$ & 6.4$^j$
\\
\hline\hline
\end{tabular}
\caption{Bounds from flavor-changing neutral currents on $M\, [\TeV]
/\sqrt{|\lambda_i \lambda_j|}$ in the leptonic models, and from the $\Delta F=1$
constraints on the hadronic models.  The $\Delta F=2$ bounds show
$M/\sqrt{|\lambda_i\lambda_j|^2}$, except for $K$ meson mixing we show $\big\{
M/\sqrt{|{\rm Re}(\lambda_i\lambda_j^*)^2|}, \ M/\sqrt{|{\rm
Im}(\lambda_i\lambda_j^*)^2|}\, \big\}$.  The strongest bounds arise from: $a)$
$\mu$ to $e$ conversion; $b)$ $\tau\to e\pi$; $c)$ $\tau\to \mu\rho$; $d)$
$K\to\pi\nu\bar\nu$; $e)$ $K_L\to\mu^+\mu^-$ (this bound involves $|{\rm
Re}(\lambda_1\lambda_2^*)|$); $f)$ $K$ mixing; $g)$ $B\to\pi\mu^+\mu^-$; $h)$
$B_d$ mixing; $i)$ $B\to X_s\ell^+\ell^-$; $j)$ $B_s$ mixing; $k)$
$D\to\mu^+\mu^-$; $l)$ $D$ mixing; $m)$ $B_s\to\mu^+\mu^-$.}
\label{tab:bounds}
\end{table*}
\endgroup

We considered the experimental constraints from flavor physics on the
mass of these vector-like fermions to get a feel for the mass reach
that the present experiments have for this class of models. We are
primarily interested in very heavy vector-like fermions, say, with
masses greater than 10~TeV. Hence it is flavor-changing neutral-current 
processes that provide the most important constraints. However, we 
also discussed violations of lepton universality and CKM matrix unitarity.

An important feature of these models is that for large $M$ the BSM
contribution to meson mixing is either suppressed by a loop factor
$(\sim 1/16 \pi^2)$ or a factor of $m_Z^2/M^2$ compared to processes
that change flavor by one unit and are dominated by tree level through
flavor-changing $Z$ exchange. This implies that, except for the case of
the kaon $CP$ violation parameter $\epsilon_K$, the constraints on $M$
from meson mixing are not overwhelmingly strong. We computed the order $1/(4 \pi M)^2$ one-loop contribution to the coefficients of the four-quark operators responsible for meson mixing that is not suppressed by SM quark masses or weak mixing angles.

We are interested in a rough assessment of the experimental reach.
The strongest current bounds on the vector-like fermion masses and
couplings in each of the 11 models studied in this paper are
summarized in Table~\ref{tab:bounds}.  We display bounds which are
above or will get near the 10~TeV level in the near future.  Muon to
electron conversion in a nucleus, $\mu\to 3e$, the kaon $CP$-violating
parameter $\epsilon_K$, and $K_L\to\mu^+\mu^-$ are sensitive to
vector-like fermion masses $\gtrsim 100\,\TeV$ (for Yukawa couplings
with magnitude unity).  However, it is important to remember that the
couplings $\lambda_i$ may have a flavor structure that suppresses
these contributions relative to those involving the third generation.

\begingroup
\begin{table*}[tb] \tabcolsep 6pt
\begin{tabular}{c@{\extracolsep{2pt}}l|cc|cc|cc}
\hline\hline
\multirow{2}{*}{Model} &  \multicolumn{1}{c|}{Quantum}  &
  \multicolumn{6}{c}{Future bounds on $M/\TeV$ and $\lambda_i \lambda_j$
  for each $ij$ pair}  \\
& \multicolumn{1}{c|}{numbers} & \multicolumn{2}{c|}{$ij=12$} 
  & \multicolumn{2}{c|}{$ij=13$} & \multicolumn{2}{c}{$ij=23$}  \\
\hline
I & $(1,1,-1)$ & \multicolumn{2}{c|}{2000$^a$} & \multicolumn{2}{c|}{19$^b$}
  & \multicolumn{2}{c}{21$^c$} \\
II & $(1,3,-1)$ & \multicolumn{2}{c|}{1400$^a$} & \multicolumn{2}{c|}{13$^b$}
  & \multicolumn{2}{c}{15$^c$}
\\
III & $(1,2,-1/2)$ & \multicolumn{2}{c|}{2000$^a$} & \multicolumn{2}{c|}{19$^b$}
  & \multicolumn{2}{c}{21$^c$}
\\
IV & $(1,2,-3/2)$ & \multicolumn{2}{c|}{2000$^a$} & \multicolumn{2}{c|}{19$^b$}
  & \multicolumn{2}{c}{21$^c$}
\\ \cline{3-8}
& & $\Delta F=1$ & $\Delta F=2$ & $\Delta F=1$ & $\Delta F=2$
  & $\Delta F=1$ & $\Delta F=2$
\\ \cline{3-8}
V & $(3,1,-1/3)$ & 280$^d$ & \{$100$, 1000\}$^e$
  & 60$^f$ & 61$^g$ & 39$^h$ & 14$^i$
\\
VI & $(3,1,2/3)$ & 8.3$^j$ & \{$100$, 1000\}$^e$ 
  & --- & 61$^g$ & --- & 14$^i$
\\
VII & $(3,3,-1/3)$  & 200$^d$ & \{$110$, 1100\}$^e$
  & 42$^f$ & 68$^g$ & 28$^h$ & 16$^i$
\\
VIII & $(3,3,2/3)$  & 280$^d$ & \{$110$, 1100\}$^e$
  & 60$^f$ & 68$^g$ & 39$^h$ & 16$^i$
\\
IX $\lambda^{(u)}$ & \multirow{2}{*}{$\bigg\}(3,2,1/6)$}
  & 8.3$^j$ & 67$^k$ & --- & 86$^g$ & --- & 20$^i$
\\
IX $\lambda^{(d)}$ &  & 280$^d$ & \{$140$, 1400\}$^e$
  & 60$^f$ & 86$^g$ & 39$^h$ & 20$^i$
\\
X & $(3,2,7/6)$ & 8.3$^j$ & 48$^k$ & --- & --- & --- & ---
\\
XI & $(3,2-5/6)$  & 280$^d$ & \{$100$, 1000\}$^e$
  & 60$^f$ & 61$^g$ & 39$^h$ & 14$^i$
\\
\hline\hline
\end{tabular}
\caption{Expected future bounds from flavor-changing neutral currents on $M\,
[\TeV] /\sqrt{|\lambda_i \lambda_j|}$ in the leptonic models, and from the
$\Delta F=1$ constraints on the hadronic models.  The $\Delta F=2$ bounds show
$M/\sqrt{|\lambda_i\lambda_j|^2}$, except for $K$ meson mixing we show $\big\{
M/\sqrt{|{\rm Re}(\lambda_i\lambda_j^*)^2|}, \ M/\sqrt{|{\rm
Im}(\lambda_i\lambda_j^*)^2|}\, \big\}$.  The bounds are from: $a)$ $\mu$ to $e$
conversion; $b)$ $\tau\to e\pi$; $c)$ $\tau\to \mu\rho$; $d)$
$K\to\pi\nu\bar\nu$; $e)$ $K$ mixing; $f)$ $B_d\to\mu^+\mu^-$; $g)$ $B_d$
mixing; $h)$ $B_s\to\mu^+\mu^-$; $i)$ $B_s$ mixing; $j)$ $D\to\mu^+\mu^-$; $k)$
$D$ mixing.}
\label{tab:future}
\end{table*}
\endgroup

We summarize the expected future sensitivity in
Table~\ref{tab:future}, where we only display sensitivities near or
above 10~TeV.  The Mu2e constraint will be improved dramatically in
the next generation of experiments~\cite{Abrams:2012er,
  Kurup:2011zza}. The measured $K^+ \to \pi^+ \nu \bar\nu$ branching
ratio corresponds to a mass reach around 70~TeV, which will increase
substantially as the next generation experiments reach an uncertainty
at about 10\% of the SM rate, especially since the current central
value is above the SM prediction (which has very small theoretical
uncertainty).  The improvement in the $\Delta m_K$ bound is entirely
dependent on lattice QCD calculations, as discussed after
Eq.~(\ref{mixboundK}).  The $B_{d,s}$ and $D$ mixing sensitivities
will be improved by Belle~II and LHCb.  These experiments, and CMS and
ATLAS, will also probe FCNC $B_{d,s}$, $D$, and $\tau$ decays much
better than current bounds.  The future sensitivities in
Table~\ref{tab:future} correspond to estimated 50/ab
Belle~II~\cite{Belle2predictions} and 50/fb
LHCb~\cite{LHCb-PUB-2014-040} sensitivities and CMS/ATLAS reach in
rare decays on the same time scale.  Compared to
Table~\ref{tab:bounds}, a greater number of the best bounds will come
from purely leptonic rather than semileptonic decays.  The
sensitivities in Table~\ref{tab:future} may be realized in $\sim10$
years.

\acknowledgments

We thank Doug Bryman, Tim Gershon, Yossi Nir, Karim Trabelsi, and
Phill Urquijo for helpful comments.  MBW thanks the Perimeter
Institute for their hospitality during the completion of this work.
ZL thanks the hospitality of the Aspen Center for Physics, supported
by the NSF Grant No.~PHY-1066293, during the completion of this work.
ZL was supported in part by the Office of Science, Office of High
Energy Physics, of the U.S.\ Department of Energy under contract
DE-AC02-05CH11231.  MBW was supported by the Gordon and Betty Moore
Foundation through Grant No.\ 776 to the Caltech Moore Center for
Theoretical Cosmology and Physics, and by the DOE Grant DE-SC0011632.
He is also grateful for the support provided by the Walter Burke
Institute for Theoretical Physics.

\appendix

\section{Diagonalizing matrix}
\label{app:DiagMatrix}

Here we summarize the $4\times 4$ diagonalizing matrices in the 11 models. 
Terms of order $(v/M)^2$ and higher are not explicitly displayed.
\begin{align}
{\rm Model~I}:~~
  &V_L^{\hat{e}}= \left(
  \begin{array}{cc}
    1 & \lambda_i v/M \\
    -\lambda_i^* v/M & 1_{3\times 3}
  \end{array}
  \right) , \quad V_R^{\hat{e}}=1_{4\times 4}\,, \\
{\rm Model~II}:~~
  &V_L^{\hat{e}}= \left(
  \begin{array}{cc}
    1 & -\lambda_i v/\sqrt{2}M \\
    \lambda_i^* v/\sqrt{2}M & 1_{3\times 3}
  \end{array}
  \right) , \quad V_R^{\hat{e}}=1_{4\times 4}\,, \\
{\rm Model~III}:~~
  &V_L^{\hat{e}}= 1_{4\times 4}\,, \quad
  V_R^{\hat{e}}=\left(
  \begin{array}{cc}
    1 & -\lambda_i^* v/M \\
    \lambda_i v/M & 1_{3\times 3}
  \end{array}
  \right) , \\
{\rm Model~IV}:~~ 
  &V_L^{\hat{e}}= 1_{4\times 4}\,, \quad
  V_R^{\hat{e}}=\left(
  \begin{array}{cc}
    1 & \lambda_i^* v/M \\
   - \lambda_i v/M & 1_{3\times 3}
  \end{array}
  \right) , \\
{\rm Model~V}:~~
  &V_L^{\hat{d}}= \left(
  \begin{array}{cc}
    1 & \lambda_i v/M \\
    -\lambda_i^* v/M & 1_{3\times 3}
  \end{array}
  \right) , \quad V_R^{\hat{d}}=1_{4\times 4}\,, \\
{\rm Model~VI}:~~
  &V_L^{\hat{u}}= \left(
  \begin{array}{cc}
    1 & -\lambda_i v/M \\
    \lambda_i^* v/M & 1_{3\times 3}
  \end{array}
  \right) , \quad V_R^{\hat{u}}=1_{4\times 4}\,, \\
{\rm Model~VII}:~~
  &V_L^{\hat{u}}= \left(
  \begin{array}{cc}
    1 & \lambda_i v/M \\
    -\lambda_i^* v/M & 1_{3\times 3}
  \end{array}
  \right) , \quad V_R^{\hat{u}}=1_{4\times 4}\,, \\
  &V_L^{\hat{d}}= \left(
  \begin{array}{cc}
    1 & -\lambda_i v/\sqrt{2}M \\
    \lambda_i^* v/\sqrt{2}M & 1_{3\times 3}
  \end{array}
  \right)\,, \quad V_R^{\hat{d}}=1_{4\times 4}\,, \\
{\rm Model~VIII}:~~
  &V_L^{\hat{u}}= \left(
  \begin{array}{cc}
    1 & -\lambda_i v/\sqrt{2}M \\
    \lambda_i^* v/\sqrt{2}M & 1_{3\times 3}
  \end{array}
  \right) , \quad V_R^{\hat{u}}=1_{4\times 4}\,, \\
  &V_L^{\hat{d}}= \left(
  \begin{array}{cc}
    1 & -\lambda_i v/M \\
    \lambda_i^* v /M & 1_{3\times 3}
  \end{array}
  \right) , \quad V_R^{\hat{d}}=1_{4\times 4}\,, \\
{\rm Model~IX}:~~
        &V_L^{\hat{u}}=1_{4\times 4}\,, \quad
        V_R^{\hat{u}}=\left(
  \begin{array}{cc}
    1 & \lambda_i^{(u)*} v/M \\
    -\lambda_i^{(u)} v/M & 1_{3\times 3}
  \end{array}
  \right) ,  \\
  &V_L^{\hat{d}}=1_{4\times 4}\,, \quad 
  V_R^{\hat{d}}=\left(
  \begin{array}{cc}
    1 & -\lambda_i^{(d)*} v/M \\
    \lambda_i^{(d)} v /M & 1_{3\times 3}
  \end{array}
  \right) ,  \\
{\rm Model~X}:~~
        &V_L^{\hat{u}}=1_{4\times 4}\,, \quad
        V_R^{\hat{u}}=\left(
  \begin{array}{cc}
    1 & -\lambda_i^{*} v/M \\
    \lambda_i v/M & 1_{3\times 3}
  \end{array}
  \right) ,  \\
{\rm Model~XI}:~~
   &V_L^{\hat{d}}=1_{4\times 4}\,, \quad
   V_R^{\hat{d}}=\left(
  \begin{array}{cc}
    1 & -\lambda_i^{*} v/M \\
    \lambda_i v/M & 1_{3\times 3}
  \end{array}
  \right) .
\end{align}

Finally, there are tree-level mass splittings among the heavy fermions, except
for the $SU(2)_L$ singlet models, due to electroweak symmetry breaking. The
results are
\begin{align}
{\rm Model~II}:~~
  &M_0 = M + \Delta M\,, \qquad M_{-1} = M+\Delta M/2\,, \qquad M_{-2} = M\,, \\
{\rm Model~III}:~~
  &M_0 = M \,, \qquad M_{-1} = M + \Delta M\,, \\
{\rm Model~IV}:~~ 
  &M_{-1} = M + \Delta M\,, \qquad M_{-2} =M , \\
{\rm Model~VII}:~~
  &M_{2/3}=M+\Delta M\,, \qquad M_{-1/3}=M+\Delta M/2\,, \qquad M_{-4/3}=M \,, \\
{\rm Model~VIII}:~~
  &M_{5/3}=M\,, \qquad M_{2/3}=M+\Delta M/2\,, \qquad M_{-1/3}=M+\Delta M\,, \\
{\rm Model~IX}:~~
  &M_{2/3} = M_{-1/3} = M + \Delta M \,, \\
{\rm Model~X}:~~
  &M_{5/3} = M\,, \qquad  M_{2/3} = M + \Delta M \,, \\
{\rm Model~XI}:~~
  &M_{-1/3} = M + \Delta M \,, \qquad M_{-4/3} = M \,.
\end{align}
In each model $\Delta M = (v^2 / 2M) \sum_i |\lambda_i|^2$.

\section{Numerical inputs and Wilson coefficients}
\label{AppB}

We collect in this Appendix expressions used to derive bounds on
vector-like fermions from flavor-changing neutral-current processes,
to facilitate easier comparison with and reproduction of our numerical
results.  Our goal in this paper is to study the sensitivities of many
processes, so leading or next-to leading order results suffice.  (For
most processes the state of the art is one or two orders higher.)  In
many cases we ignore the SM uncertainties, when we know that they are
subdominant effects.

We adopt for the numerical values of the coupling constants
$\alpha(m_b) = 1/133$, $\sin^2\theta_W = 0.23$, $g_Z = 0.73$.  For the
top quark mass we use $\overline m_t(m_t) = 165\,\GeV$, obtained from
the one-loop relation from the $m_t = 173\,\GeV$ (presumed) pole mass,
extracted from fits to $t\bar t$ production at the Tevatron and the
LHC.

The SM Wilson coefficients are as follows.  For $K^+\to
\pi^+\nu_i\bar\nu_i$~\cite{Buchalla:1993wq,Buchalla:1998ba,
  Buras:2004uu,Brod:2008ss,Buras:2015qea}
\begin{equation}
c_{+\nu}^{\rm (SM)} = c_\nu^{\rm (SM)} = 
  \frac{\sqrt2\, G_F\, \alpha}{\pi\, \sin^2\theta_W}\,
  \Big[V_{ts}^*V_{td} X(x_t) + V_{cs}^*V_{cd} \lambda^4 P_c(X) \Big]\, ,
\end{equation}
where $x_t=\overline m_t^2/m_W^2$, $\lambda=0.225$, $P_c(X) \simeq
0.4$~\cite{Buras:2004uu,Buras:2015qea},\footnote{In our definition of
  $c_\nu^{\rm (SM)}$ all lepton flavors (labeled by $i$) are included
  by using parameter $P_c(X)$, which is defined in
  \cite{Buchalla:1993wq,Buchalla:1998ba}. Then $c_\nu^{\rm (SM)}$
  corresponds to the decay amplitude.} and
\begin{equation}
  X(x_t) = \frac{x_t}{8}
  \left[\frac{x_t+2}{x_t-1}+\frac{(3x_t-6)\ln x_t}{(x_t-1)^2} \right]\, .
\end{equation}
Here we have neglected the electroweak corrections.

For $B_q\to \mu^+\mu^-$ (where $q=s,d$), 
\begin{equation}
c_{10}^{\rm (SM)} = \frac{G_F\, \alpha}{\sqrt2\, \pi}\,
  V_{tb}^*V_{tq}\, \frac{Y(x_t)}{\sin^2\theta_W}\,, \qquad
Y(x_t) = \frac{x_t}8 \left(\frac{x_t-4}{x_t-1} + \frac{3x_t\ln
x_t}{(x_t-1)^2} \right) .
\end{equation}
(In the usual notation, $C_{10} = -Y(x_t) / \sin^2\theta_W \simeq -4.2$.)

The short-distance contribution to the $K_L \to \mu^+\mu^-$ rate
can be written as~\cite{Buras:2013rqa} 
\begin{equation}
\Gamma(K_L \to \mu^+\mu^-)_{\rm SD} = 
  \frac{G_F^4 m_W^4}{4\pi^5} \sin^4\theta_W
  f_K^2 m_K m_\mu^2 \sqrt{1-\frac{4m_\mu^2}{m_K^2}}\,
  {\rm Re}\Big[\Big(V_{ts}^*V_{td}\, Y(x_t) + V_{cs}^*V_{cd}\, \lambda^4 P_c
  \Big)^{\!2}\,\Big] ,
\end{equation}
where $P_c \simeq 0.11$.

The $B\to K \mu^+\mu^-$ rate is given by~\cite{Buchalla:2000sk}
\begin{align}\label{BtoKrate}
\frac{{\rm d}\Gamma(B\to K \mu^+\mu^-)}{{\rm d}q^2} = &
  \frac{G_F^2\alpha^2m_B^3}{1536\pi^5} |V_{tb}V_{ts}|^2\, [\lambda_K(q^2)]^{3/2}
  \bigg\{ f_+^2\, \big[ |C_9^{\rm eff}(q^2)|^2 + |C_{10} + C'_{10}|^2 \big] 
  \nonumber\\
& + \frac{4m_b^2}{(m_B+m_K)^2}\, f_T^2\, |C_7|^2 +
  \frac{4m_b}{m_B+m_K}\, f_T f_+\, {\rm Re} \big[C_9^{\rm eff}(q^2) C_7^*\big]
  \bigg\} \,,
\end{align}
where $\lambda_K(q^2)$ is a phase space factor, $f_+$ and $f_T$ are
$q^2$-dependent form factors.  The $B\to \pi\mu^+\mu^-$ rate is
obtained with obvious replacements. In the heavy quark limit $f_+ /
f_T = 1 +{\cal O}(\Lambda_{\rm QCD}/m_b)$~\cite{Isgur:1990kf}, and
model calculations are consistent with a mild $q^2$ dependence of this
ratio.  Hence, at the desired level of precision, the form factors can
be pulled out of the $\{ \ldots \}$ expression in
Eq.~(\ref{BtoKrate}), obtaining simple approximations for the effect
of new physics via $C_{10}$ and $C'_{10}$.  (This is impossible for
$B\to K^* \mu^+\mu^-$, as $C_7$ enters with a $1/q^2$ dependence in
that case.)  In the numerical analysis, we use $C_7=-0.33$,
$C_{10}=-4.2$, and a mean value $|C_9^{\rm eff}(q^2)| = 4.4$.  These
estimates can be refined as the measurements improve.

\end{document}